\documentclass[aip,amsmath,amssymb,floatfix,reprint]{revtex4-1}
\pdfoutput=1

\usepackage[utf8]{inputenc}
\usepackage{graphicx}
\usepackage{color}
\usepackage{soul}
\usepackage{gensymb}    
\usepackage{textcomp}   

\usepackage{pdfpages}

\makeatletter
\AtBeginDocument{\let\LS@rot\@undefined}
\makeatother

\graphicspath{{figures/}}

\newcommand{\wave}{cm$^{-1}$}
\newcommand{\appr}{$\sim$}

\begin{document}

\def\mytitle{The Interplay of Structure and Dynamics in the Raman Spectrum of Liquid Water over the Full Frequency and Temperature Range}
\title{\mytitle}

\author{Tobias Morawietz}
\affiliation{
Department of Chemistry, Stanford University, Stanford, CA 94305, United States
}

\author{Ondrej Marsalek}
\affiliation{
Faculty of Mathematics and Physics, Charles University, Ke Karlovu 3, 121 16 Prague 2, Czech Republic
}

\author{Shannon R. Pattenaude}
\affiliation{
Department of Chemistry, Purdue University, West Lafayette, IN 47907, United States
}

\author{Louis M. Streacker}
\affiliation{
Department of Chemistry, Purdue University, West Lafayette, IN 47907, United States
}

\author{Dor Ben-Amotz}
\affiliation{
Department of Chemistry, Purdue University, West Lafayette, IN 47907, United States
}

\author{Thomas E. Markland}
\email{tmarkland@stanford.edu}
\affiliation{
Department of Chemistry, Stanford University, Stanford, CA 94305, United States
}

\date{\today}

\begin{abstract}
While many vibrational Raman spectroscopy studies of liquid water have investigated the
temperature dependence of the high-frequency O-H stretching region, few have analyzed the
changes in the Raman spectrum as a function of temperature over the entire spectral range. Here,
we obtain the Raman spectra of water from its melting to boiling point, both experimentally and
from simulations using an ab initio-trained machine learning potential. We use these to assign
the Raman bands and show that the entire spectrum can be well described as a combination of
two temperature-independent spectra. We then assess which spectral regions exhibit strong
dependence on the local tetrahedral order in the liquid. Further, this work demonstrates that
changes in this structural parameter can be used to elucidate the temperature dependence of the
Raman spectrum of liquid water and provides a guide to the Raman features that signal water
ordering in more complex aqueous systems.
\end{abstract}

{\maketitle}

\section*{TOC graphic}

\begin{center}
  \includegraphics{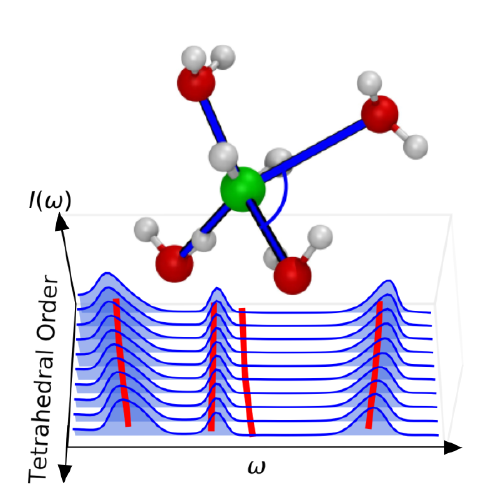}
\end{center}

\section{Introduction}

Despite the importance of liquid water and the structural and dynamic sensitivity of vibrational
Raman spectroscopy, quantitatively linking water spectra and structure remains a challenge for
experiment and theory.\cite{Bakker2010} Indeed, few experimental studies have spanned the entire liquid
temperature and vibrational frequency range: from the low frequency intermolecular hydrogen
bond (H-bond) stretch, O-H...O at \appr180 \wave , to the high-frequency O-H stretch region at $\sim$3400
\wave.\cite{Scherer1974,Zhelyaskov1989,Bray2013,DeSantis1987,Walrafen1973,Murphy1972,Hare1990,Walrafen1986} 
The need for such experimental results is particularly timely as it is becoming increasingly practical to 
simulate the Raman spectra of water using first principles approaches across the entire frequency range,\cite{Wan2013a,Medders2015,Marsalek2017} 
thus providing a strenuous test of these methods' ability to correctly capture and elucidate the structure 
and dynamics of water. Here we employ a combined experimental and theoretical strategy to address the open 
questions regarding the origin of vibrational features in the Raman spectra of liquid water from its melting 
to boiling point.

Many previous experimental Raman studies\cite{Scherer1974,Hare1990,Walrafen1986,DArrigo1981,Walrafen1986a,Sun2013} have concentrated on analyzing the
temperature dependence of the O-H stretching region where an isosbestic point, a region in the
spectrum where the intensity is approximately constant upon a change in temperature,\cite{Walrafen1986a} is
observed. The bimodal profile of the isotropic line shape together with the observation of an
isosbestic point has been frequently attributed to an equilibrium between O-H bonds that
correspond to water molecules in two different local environments.\cite{Walrafen1986,DArrigo1981,Sun2013,Harada2017} Although such
isosbestic behavior is expected for spectra composed of two components, it can also arise from a
continuous distribution of thermally equilibrated structures.\cite{Geissler2005,Smith2005} Theoretical studies have thus
sought to simulate and decompose the Raman spectra. For example the temperature dependence
of the isotropic Raman O-H stretching band has been shown to be remarkably well reproduced
by simulations employing rigid water models using mappings between the vibrational frequency
and the local electric field\cite{Corcelli2004,Corcelli2005,Auer2008,Yang2010,Tainter2013} while some early studies have calculated the low frequency
terahertz region from time-correlation functions of the polarizability tensor.\cite{Madden1986,Mazzacurati1989,Bursulaya1998} However, more
recently it has become possible to use high-level ab initio-based potential energy surfaces\cite{Medders2015} or ab
initio molecular dynamics (AIMD) calculations with classical\cite{Wan2013a,Marsalek2017} and even quantum nuclei\cite{Marsalek2017} to
make fully first principles predictions of the Raman spectrum at ambient conditions across the
entire frequency range.

Here we present a combined experimental and theoretical study of the temperature-
dependent Raman spectra of liquid water from its freezing to boiling point over the full
frequency range, from \appr100 \wave\,to \appr4200 \wave. By doing this we address open questions
regarding the origin of vibrational features that are more prominent in the Raman than Infrared
(IR) spectra and the correlation between the vibrational and structural properties of water. To
increase the efficiency of our simulations we employ neural network potentials (NNPs)\cite{Behler2007,Artrith2016,Behler2016,Morawietz2016}
trained to density functional theory calculations (see Supporting Information (SI), sections 1-3).
By experimentally and theoretically probing the entire vibrational frequency range here we
provide a rigorous assignment of the low-intensity modes in the vibrational spectra and identify
several spectral regions, in addition to the O-H stretching region, that exhibit strong dependence
on the local tetrahedral order of the liquid. Our results further reveal that the temperature
dependence of both the vibrational spectrum of water and its tetrahedral order distribution can be
accurately decomposed into a linear combination of two temperature-independent components.
By employing a time-dependent analysis of our simulated spectra, we provide theoretical support
for the empirical observation that enhanced tetrahedral order is associated with features
appearing across the entire frequency range, from the low-frequency H-bond stretch band to the
high-frequency O-H stretch band. This analysis allows us to identify the regions that provide the
most sensitive spectral signatures of structural ordering in liquid water, thus offering insights into
the origins of these features. The identification of these features will aid in the analysis of other
complex aqueous environments ranging from the hydration-shells of solute molecules to
catalytic surfaces and biological interfaces.\cite{Heyden2008,Artrith2014,Gierszal2011,Fayer2010,Crans2012,Davis2012a,Bakulin2013,Russo2017}

\section{Results and Discussion}

\begin{figure}[htbp]
\centering
  \includegraphics{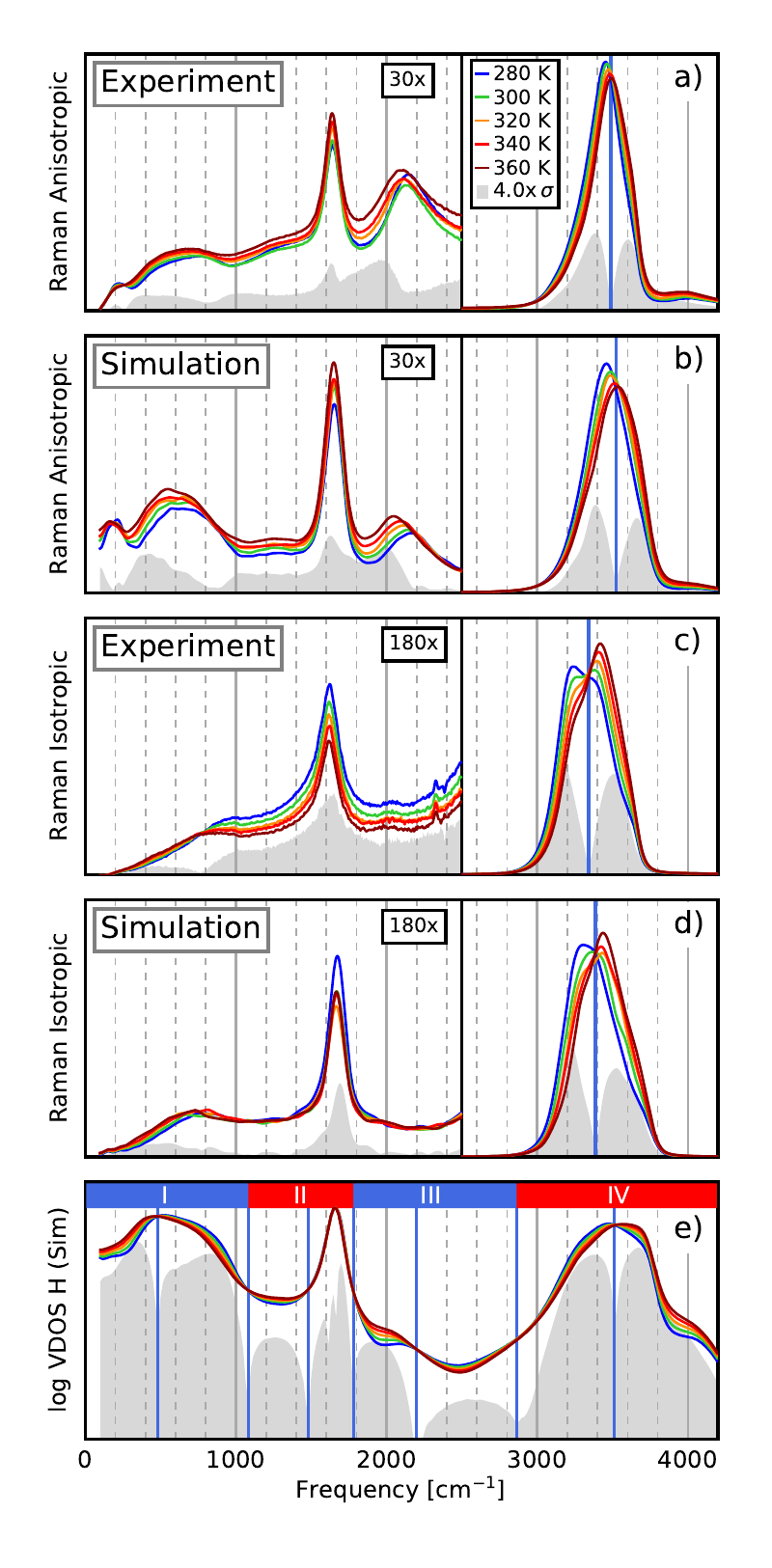}\\  
\caption{
\label{fig:fig01}
Comparison of the experimental and simulated anisotropic and isotropic reduced
Raman spectra (a-d) and the simulated hydrogen vibrational density of states (VDOS, shown on
a logarithmic scale to highlight the low-intensity features) (e) of liquid water at different
temperatures. For clarity, the low frequency region was scaled by the factor indicated in the plot.
All spectra are truncated at 100 \wave\,which is the lowest frequency for which accurate 
experimental intensities were obtained. All spectra are normalized to unit area. The standard
deviation $\sigma$, with respect to temperature (shown as grey shading), was used to locate isosbestic
points which are depicted as blue lines. The hydrogen VDOS spectrum can be divided into four
regions, each of which has an isosbestic point at the center and is separated by isosbestic points
from the other regions.
}
\end{figure}

In Fig.~\ref{fig:fig01} we compare the experimental (panels a and c) and simulated (panels b and d)
anisotropic and isotropic Raman spectra of liquid water. The spectra are presented in a reduced
form\cite{Brooker1988,Brooker1989} (see SI sections 4-5) that is particularly useful for such a comparison, as it highlights
features in the low frequency region, and removes the dependence on the incident laser
frequency. Fig.~\ref{fig:fig01}e shows the temperature-dependent vibrational density of states (VDOS) of
the hydrogen atoms (on a logarithmic-scale for better visibility of the low-intensity regions)
which exhibits analogous trends to those seen in the Raman spectra. The low-frequency band at
\appr200 \wave, which has been attributed to the (intermolecular) H-bond stretching mode,\cite{Walrafen1986} is barely
visible in the hydrogen VDOS, but is more prominent in the oxygen VDOS (see SI section 6).
The grey shading in Fig.~\ref{fig:fig01} shows the standard deviation of the spectral data obtained at
different temperatures and thus gives an indication of the temperature sensitivity of different
spectral regions. The minima in the standard deviation thus allow for the approximate
identification of isosbestic points (frequencies where the spectral intensity is temperature-
independent).

\begin{figure}[htbp]
\centering
  \includegraphics{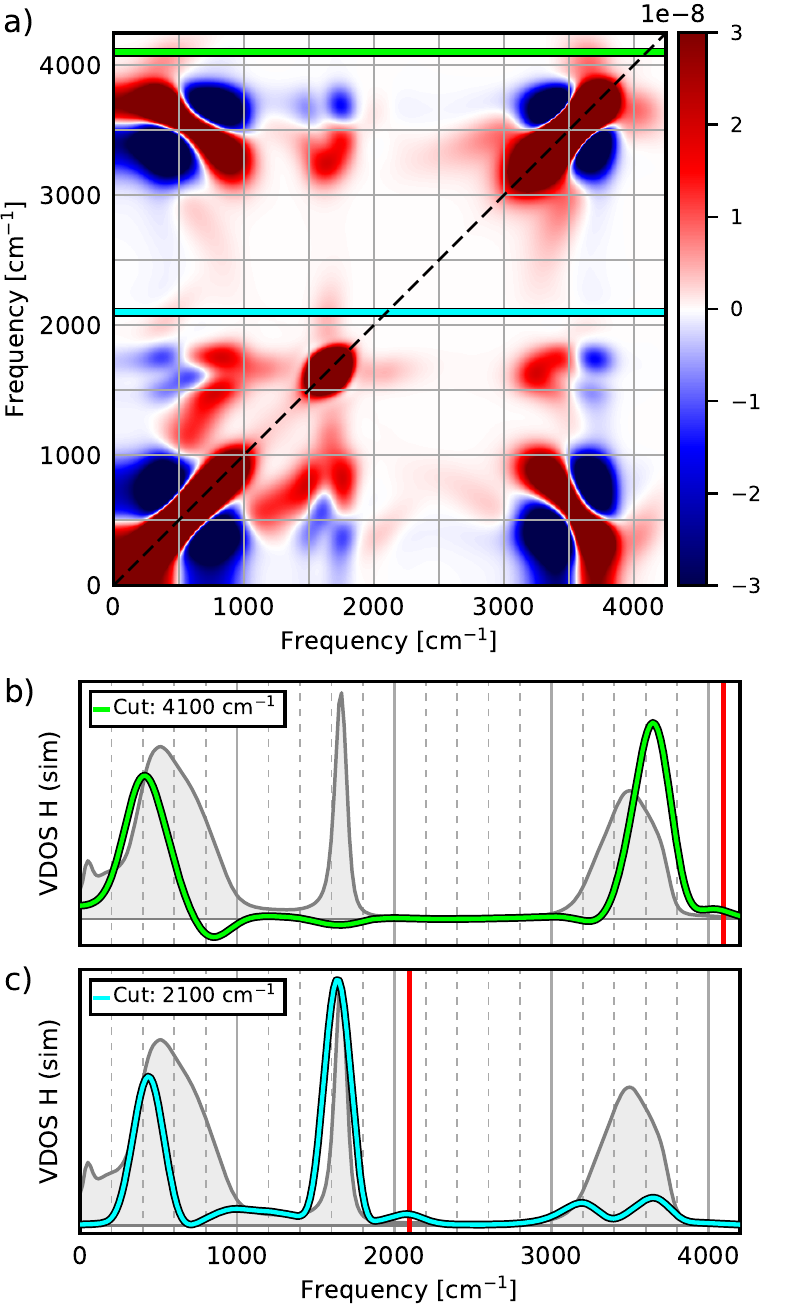}\\  
\caption{
\label{fig:fig02}
Synchronous two-dimensional correlation spectrum of the simulated hydrogen
vibrational density of states (VDOS) at T = 300 K (a). Off-diagonal regions colored in red
indicate pairs of frequencies that are positively correlated (intensity of both frequencies in the
pair change in the same direction) whereas blue regions indicate an anticorrelation (intensity of
both frequencies in the pair change in the opposite direction). Cuts at frequencies of 4100 \wave\,
and 2100 \wave, indicated by the light green and light blue horizontal lines in panel (a), are shown
in panels (b) and (c). As reference, the full hydrogen VDOS spectrum is shown in grey.
}
\end{figure}

The high-frequency O-H stretching region (from \appr2500 \wave\,to \appr4200 \wave) of the
anisotropic and isotropic Raman spectra has been the focus of the majority of previous
studies.\cite{Scherer1974,Hare1990,Walrafen1986,DArrigo1981,Walrafen1986a,Sun2013,Harada2017,Corcelli2004,Corcelli2005,Auer2008,Tainter2013} 
As observed in Fig.~\ref{fig:fig01}a, the anisotropic O-H band consists of a single
peak that decreases in intensity and shifts to higher frequencies as the temperature is raised,
which resembles the behavior seen in the IR spectrum of liquid water,\cite{Marechal2011} while the isotropic band
has a bimodal profile (Fig.~\ref{fig:fig01}c). A feature that is not evident in the IR spectrum but appears in
the anisotropic Raman spectrum is the high-frequency band at around 4000 \wave, which is higher
than even the O-H stretch frequency of the isolated molecule (\appr3750 \wave). To identify the origin
of this feature, which is observed both in our experiments and simulations, we show the
synchronous two-dimensional (2D) correlation spectrum\cite{Noda1993} of the simulated VDOS in Fig.~\ref{fig:fig02}a
(also see SI section 7). This analysis indicates the frequencies that are positively correlated (red
regions) and those that are anticorrelated (blue regions) and thus allows us to identify the
vibrational modes with which the high-frequency feature is correlated. As seen in Fig.~\ref{fig:fig02}b, the
4100 \wave\,band is strongly correlated with two bands at lower frequencies -- one with its
maximum in the libration region (centered at 433 \wave) and one in the O-H stretch region
(centered at 3645 \wave) -- whose sum yields a combined frequency of 4078 \wave. The fact that
this mode is strongly associated with two bands that sum to give its frequency supports an
assignment of this feature as a combination band (i.e. arising from anharmonic couplings of two
or more fundamental modes at frequencies slightly lower than the sum of the fundamental
frequencies). This is in contrast with a previous assignment\cite{Walrafen2004} that, while identifying the 4100 \wave\,band 
as arising from librational-vibrational coupling, assigned it as a combination of a higher
frequency librational band (730 \wave) with the low frequency part of the stretch (3423 \wave),
whereas our simulations suggest it arises from coupling between a higher frequency part of the
O-H stretch and a lower frequency librational band.

In the low-frequency region of the spectrum, we observe a broad feature centered at
~2100 \wave\,(sometimes referred to as the water association band), which is visible in both the
experimental and calculated anisotropic Raman spectra. Interestingly, this band at 2100 \wave\,was
not present in recent anisotropic Raman spectra calculated using the ab initio-based MB-pol
model. 11 Experimental and theoretical studies of the IR spectra of liquid water, ice, and
trehalose-water systems have previously assigned this feature to a combination band of the
bending mode with a librational mode\cite{Marechal1991,McCoy2014} or, alternatively, to the second overtone of a librational
mode.\cite{Devlin2001} Our 2D correlation analysis of our NNP simulations (see Fig.~\ref{fig:fig02}c) shows that the 2100
\wave\,band is a combination band of the low-frequency libration band (centered at 433 \wave) with
the bend vibration (centered at 1631 \wave) summing to a frequency of 2064 \wave.

The agreement between our measured and simulated Raman spectra in the O-H stretching
region is markedly better than that observed in another recent AIMD simulation, using a
different exchange-correlation functional, where the O-H stretching band was red-shifted by
~200 \wave and significantly broadened compared to the experiment.\cite{Wan2013a} The location of the
isosbestic points in the O-H stretch region are also well reproduced by our simulations (3532
\wave\,for the anisotropic spectrum and 3385 \wave\,for the isotropic one compared to 3490 \wave\,and
3330 \wave\,in the experiment). While the overall shape of the simulated anisotropic spectra in the
low frequency region deviates slightly from the measured spectra, the position and shape of the
individual spectral features and their variation with temperature closely match those seen in the
experiment. Having confirmed the agreement between our ab initio-based NNP simulations and
the experimental Raman spectra over the full liquid temperature range, we can now use the
simulations to relate the observed spectral features to the structural environments in the liquid.

What are the structural changes that lead to the temperature dependence of the different
vibrational features occurring across the full frequency range of the Raman spectra? To begin to
investigate this question, we first performed a self-modeling curve resolution (SMCR)\cite{Lawton1971,Tauler1995,Jiang2004}
decomposition of the temperature-dependent vibrational spectra obtained from experiment and
simulation. SMCR provides a means of decomposing a collection of two or more spectra into a
linear combination of different spectral components, each of which has exclusively positive
intensity. For example, SMCR has been used to separate aqueous solution spectra into a linear
combination of bulk water and a solute-correlated component to reveal features arising from
water molecules that are perturbed by solutes, including ions,\cite{Daly2017,Rankin2013} gases,\cite{Zukowski2017} alcohols,\cite{Davis2012a,Rankin2015,Mochizuki2016} 
aromatics,\cite{Gierszal2011,Scheu2014} surfactants,\cite{Long2015,Pattenaude2016} and polymers.\cite{Mochizuki2017} Here, we employ an SMCR analysis to assess
how accurately the temperature-dependent vibrational spectra can be approximated by a linear
combination of two components, whose relative populations, but not spectral shapes, change
with temperature.

\begin{figure}[htbp]
\centering
  \includegraphics{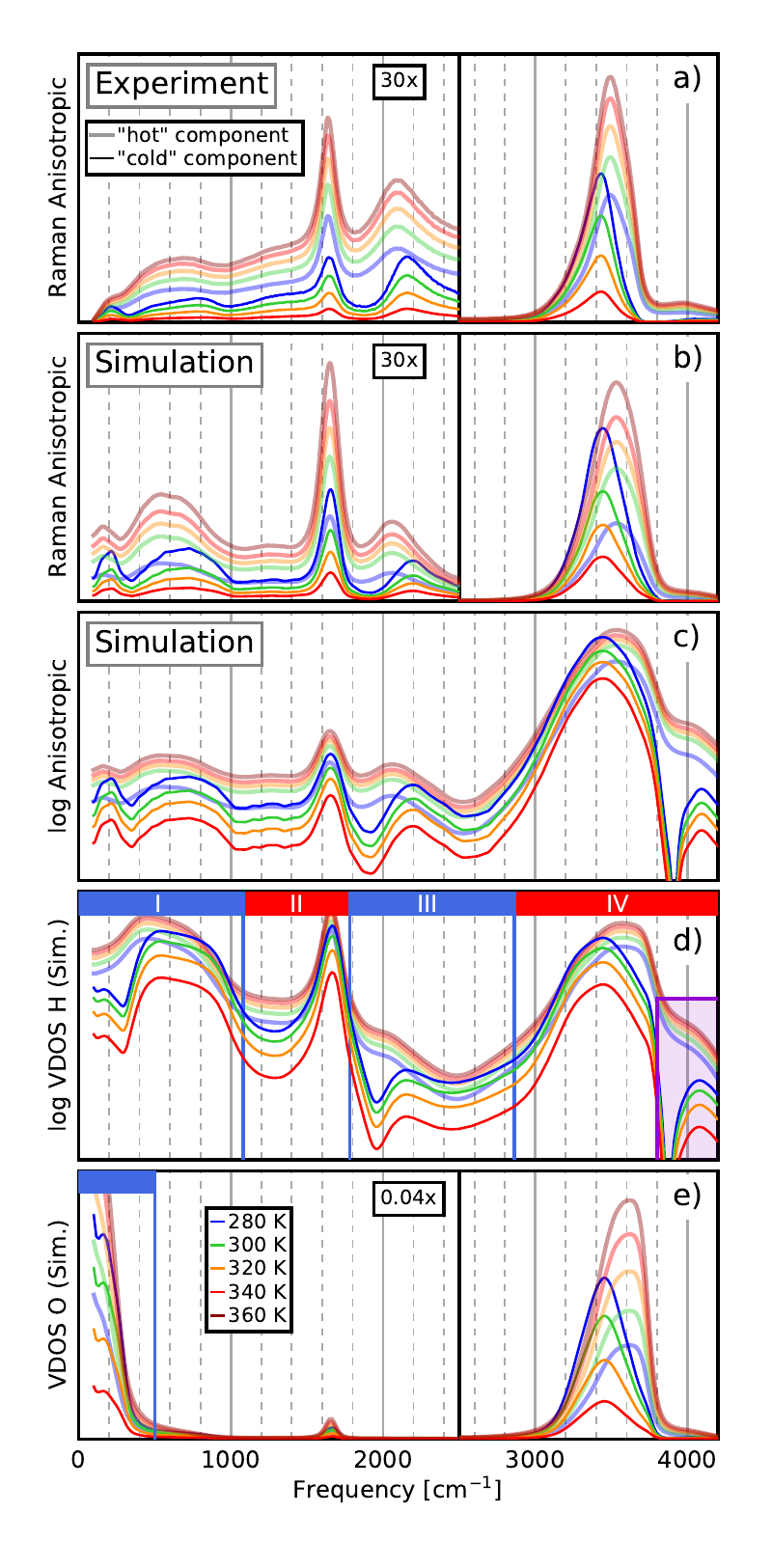}\\  
\caption{
\label{fig:fig03}
Self-modeling curve resolution (SMCR) decomposition of the experimental and
simulated anisotropic Raman spectra (a-c), and the simulated VDOS spectra for the oxygen and
hydrogen atoms (d-e). To highlight the close similarities between the anisotropic Raman and the
hydrogen VDOS spectra, both are plotted on a logarithmic scale in panels (c) and (d). The four
spectral regions in the hydrogen VDOS and an additional low-frequency region in the oxygen
VDOS are marked by blue and red bars. The purple area in the hydrogen VDOS (from 3800 \wave\,
to 4200 \wave) indicates the region that was selected for the instantaneous frequency analysis of
the high-frequency band (shown in Figure 5)
}
\end{figure}

Fig.~\ref{fig:fig03} shows the SMCR decomposition of our experimental and simulated anisotropic
Raman spectra as well as the simulated VDOS spectra, which yields a high-temperature “hot”
component, closely resembling the full spectrum at 360 K, and a “cold” component, whose
intensity increases at low temperatures and whose spectral features are shifted relative to the
“hot” component. As shown in SI Figs. S6 and S7, the reconstructed spectra, obtained by
combining the two temperature-independent “hot” and “cold” components weighted by their
populations, are almost indistinguishable from the original spectra and exhibit integrated
fractional errors below 0.01 (see SI section 8) suggesting that the vibrational spectra of water can
be accurately represented as a linear combination of two temperature-independent components
over its entire liquid temperature range.

By inspecting the two components we observe that, for the “cold” component, the
libration band and the combination band at 2100 \wave\,are shifted to higher frequencies, and that a
high-frequency peak centered at \appr4100 \wave\,appears. The most prominent bands in the low
temperature SMCR-spectra are peaked near \appr200 \wave\,and ~3400 \wave\,and resemble those
observed in ice and solid clathrate hydrates. Specifically, H$_2$O ice contains prominent bands at
\appr180 \wave\,and \appr3100 \wave.\cite{Davis2012a,Kanno1998} Similarly, the Raman spectra of various H$_2$O clathrate hydrates
contain bands peaked near \appr210 \wave\,and \appr3100 \wave.\cite{Takasu2003,Sugahara2005,Chazallon2007} The similar positions of the bands in
these tetrahedrally ordered phases to those seen in the SMCR decomposition of the liquid water
spectra suggests that these bands may provide spectroscopic probes of the local tetrahedral order
in the liquid. We note that even though the band at \appr200 \wave\,is barely visible in the SMCR
decomposed hydrogen VDOS, it can be seen much more prominently in the oxygen VDOS. This
behavior is in-line with isotope substitution studies that find larger isotope shifts of this band in
the Raman spectrum for a $^{16}$O/$^{18}$O substitution compared to a H/D substitution, suggesting that it
arises primarily from oxygen motions, consistent with the assignment of this band to the O-H...O H-bond stretch vibration.\cite{Brooker1989}

\begin{figure}[t]
\centering
  \includegraphics{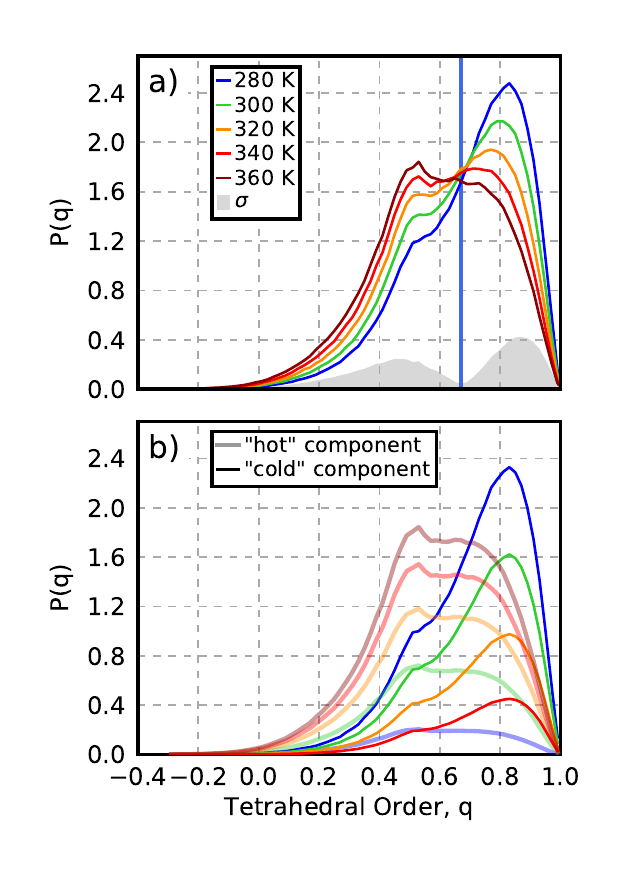}\\  
\caption{
\label{fig:fig04}
(a) Distribution of the tetrahedral order parameter, $q$, obtained from ab initio-based
simulations of liquid water as function of temperature. An isosbestic point at $q \approx$ 0.67 can be
identified (shown as blue line). (b) Self-modeling curve resolution (SMCR) decomposition of the
tetrahedral order parameter distribution.
}
\end{figure}

Given the good agreement with the experiment, we now use our simulations to assess
whether the tetrahedral order in the liquid is indeed the cause of the observed spectral shifts by
analyzing the temperature dependence of the local tetrahedral order parameter\cite{Chau1998} in its rescaled
version 64 (such that it gives a value of 1 for a regular tetrahedron and averages to 0 for an ideal
gas). The local tetrahedral order parameter (or tetrahedrality), $q_i$, is a measure for the local
angular order of water molecule i, based on its four nearest neighbors. The tetrahedrality
distribution computed from our simulations (Fig.~\ref{fig:fig04}a), shows the bimodal structure seen in
many previous simulations of liquid water,\cite{Russo2017,Errington2001a,Paolantoni2009} with an isosbestic point at $q \approx$ 0.67. Upon
cooling the distribution shifts to the right, indicating the more predominantly tetrahedral
character expected at lower temperatures. To relate the vibrational spectra to the tetrahedrality of
the liquid we perform an SMCR decomposition of the tetrahedral order parameter, shown in
Fig.~\ref{fig:fig04}b, analogous to the decomposition of the vibrational spectra. From this we see that, like
the spectra, the tetrahedral order distribution can be accurately decomposed into two
temperature-independent components. The low-temperature component is shifted to high values
of q and the broader high-temperature component is centered at lower q values. While these
results demonstrate that the vibrational shifts and the tetrahedrality of the liquid exhibit similar
temperature dependence, they alone do not provide direct proof that the spectral shifts are caused
by the tetrahedrality of the environment.

\begin{figure*}[htbp]
\centering
  \includegraphics{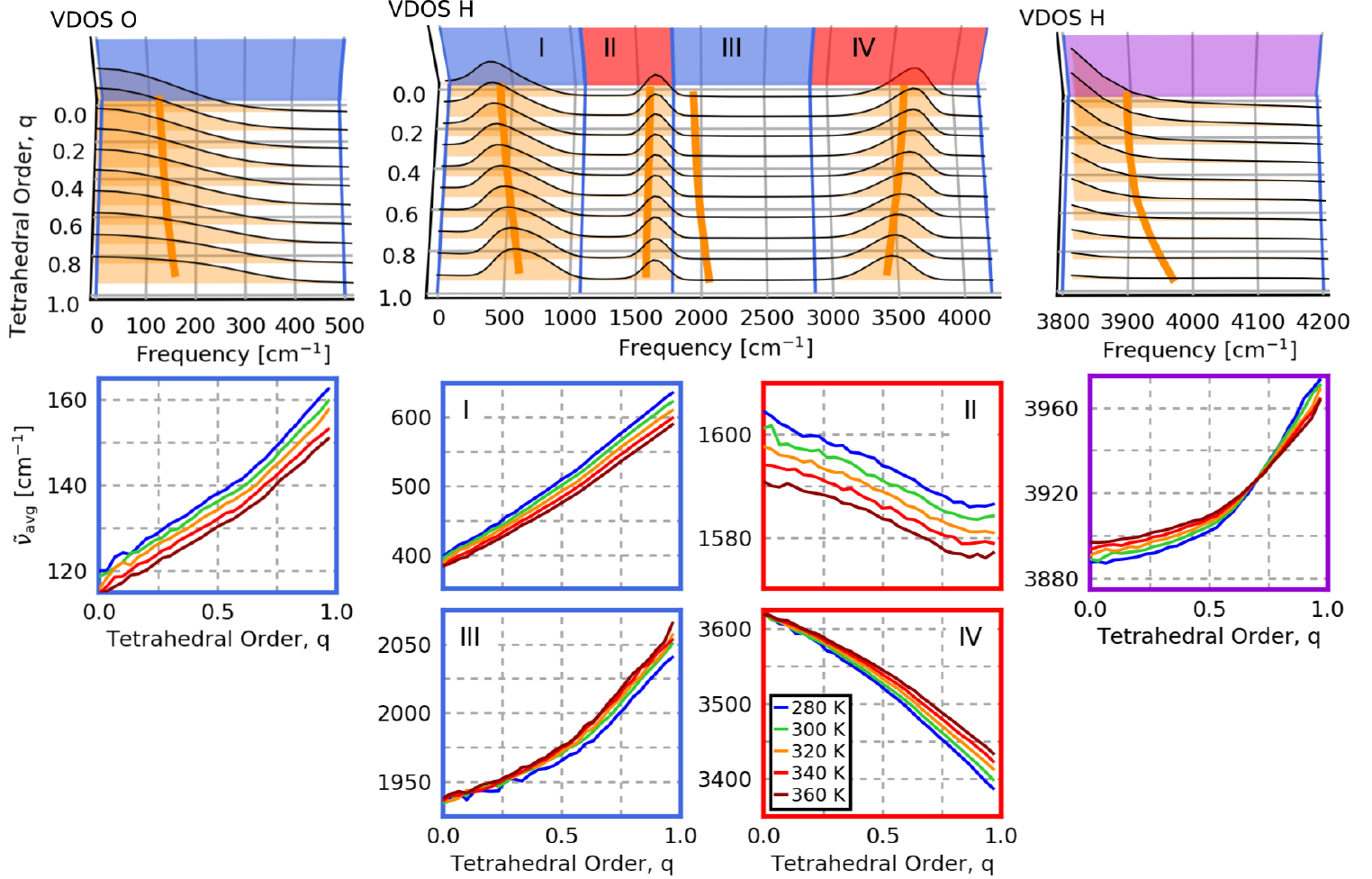}\\  
\caption{
\label{fig:fig05}
Instantaneous frequency analysis of the simulated hydrogen and oxygen VDOS spectra
as function of the tetrahedral order parameter, $q$. The top panels show the VDOS spectrum at a
single temperature (T = 320 K), decomposed into contributions from molecules with a given
tetrahedral order in the liquid water simulation. The orange lines indicate the average value of
the spectrum within the specified region. The bottom panels show the correlation of the average
peak values with q for the six different regions and all temperatures.
}
\end{figure*}

To establish the explicit connection between the spectral contribution of a water molecule
in the liquid and its tetrahedral order we thus performed a time-dependent analysis of our
simulated spectra.\cite{Marsalek2017,Napoli2007} These results are shown in Fig.~\ref{fig:fig05}, where the contribution of a given
molecule to the vibrational spectrum at a specific point in time is correlated with the
instantaneous tetrahedrality of the local hydrogen environment of the same water molecule. For
this analysis we employ the VDOS to extract the vibrational motions of individual atoms. The
top panels in Fig.~\ref{fig:fig02} show results obtained from the VDOS spectra at 320 K, binned as a
function of the tetrahedrality parameter over the full frequency range. The bottom panels show
how the average frequency in six regions of the spectrum changes with the local tetrahedral order
of the water molecule. Four of these spectral regions were defined using the seven isosbestic
points, identified from the temperature-dependent VDOS (regions I – IV, defined in Fig.~\ref{fig:fig01}e).
The other two were chosen to be the 4000 \wave combination band identified earlier (defined in
Figure 3d) and the 200 \wave O-H...O H-bond stretch vibration in the oxygen VDOS (Fig.~\ref{fig:fig03}e).

The results in Fig.~\ref{fig:fig05} (bottom panels) demonstrate that there is a direct correlation
between vibrational frequency and tetrahedrality in all spectral regions: in some regions the
spectrum shifts to higher frequencies as the tetrahedrality increases while in others it shifts to
lower frequencies. The O-H stretching band (region IV) is inversely correlated with the
tetrahedral order, while the low frequency H-bond stretching band (oxygen VDOS), the libration
band (region I), the combination band at 2100 \wave\,(region III), and the high-frequency
combination band near 4000 \wave, all are positively correlated. We observe that the bend (region
II) displays the least sensitivity with regards to tetrahedral order. The direction of all spectral
shifts with tetrahedrality is consistent with the locations of these features in the SMCR
component spectra. For instance, in the O-H stretching region the “cold” component, which is
associated with high tetrahedrality, is shifted to lower frequencies, which follows the trend seen
in the correlation plot (Fig.~\ref{fig:fig05}, lower panel, region IV) where a shift to lower frequency values
is observed as the tetrahedrality increases.

The similarity of the curves pertaining to different temperature shown in the lower panels
of Fig.~\ref{fig:fig05} further reveal that the correlation of the average peak frequency with the
tetrahedrality is relatively insensitive to temperature. This implies that the correlation between
tetrahedrality ($q$) and spectral frequency observed at a single temperature is sufficient to
approximately reconstruct the average peak positions at any temperature, given the $q$ distribution
at that temperature. This frequency-structure correlation analysis, combined with the other
results presented above, provide compelling evidence that the temperature dependence of the
Raman spectrum of water is in fact largely correlated to a single structural parameter: the
tetrahedrality of the liquid.

In summary, we have presented experiments and simulations of the temperature- and
polarization-dependent Raman spectra of liquid water over the entire vibrational frequency
range. We have shown that ab initio simulations, accelerated by machine learning potentials, are
able to accurately capture subtle temperature-dependent changes in the Raman spectrum of water
and employed a 2D correlation analysis of the simulated spectra to assign the Raman bands.
Subsequently, by linking the vibrational motions of water to the time-dependent structural
features, we have demonstrated that a single structural parameter, the local tetrahedrality, is
sufficient to predict the temperature dependence of the vibrational spectrum of liquid water
across the whole frequency range. This analysis has enabled us to identify several spectral
regions that are strongly correlated with tetrahedral order and thus could be employed in future
studies to probe the structural order of water molecules surrounding various solutes and confined
within more complex environments.

\begin{acknowledgments}
This material is based upon work supported by the National Science Foundation under Grant No.~CHE-1652960 
to T.E.M. and Grant No.~CHE-1464904 to D.B.A. T.E.M. also acknowledges
support from a Cottrell Scholarship from the Research Corporation for Science Advancement
and the Camille Dreyfus Teacher-Scholar Awards Program. T.M. is grateful for financial support
by the DFG (MO 3177/1-1). We thank Andrés Montoya-Castillo for useful discussions and
Aaron Urbas of the Bioassay Methods Group at the National Institute of Standards and
Technology for allowing us to borrow the NIST SRM$^{\text{\textregistered}}$ 2243 used for this study.
\end{acknowledgments}

\newpage

\bibliographystyle{apsrev4-1}
\bibliography{mendeley} 

\begin{thebibliography}{66}%
\makeatletter
\providecommand \@ifxundefined [1]{%
 \@ifx{#1\undefined}
}%
\providecommand \@ifnum [1]{%
 \ifnum #1\expandafter \@firstoftwo
 \else \expandafter \@secondoftwo
 \fi
}%
\providecommand \@ifx [1]{%
 \ifx #1\expandafter \@firstoftwo
 \else \expandafter \@secondoftwo
 \fi
}%
\providecommand \natexlab [1]{#1}%
\providecommand \enquote  [1]{``#1''}%
\providecommand \bibnamefont  [1]{#1}%
\providecommand \bibfnamefont [1]{#1}%
\providecommand \citenamefont [1]{#1}%
\providecommand \href@noop [0]{\@secondoftwo}%
\providecommand \href [0]{\begingroup \@sanitize@url \@href}%
\providecommand \@href[1]{\@@startlink{#1}\@@href}%
\providecommand \@@href[1]{\endgroup#1\@@endlink}%
\providecommand \@sanitize@url [0]{\catcode `\\12\catcode `\$12\catcode
  `\&12\catcode `\#12\catcode `\^12\catcode `\_12\catcode `\%12\relax}%
\providecommand \@@startlink[1]{}%
\providecommand \@@endlink[0]{}%
\providecommand \url  [0]{\begingroup\@sanitize@url \@url }%
\providecommand \@url [1]{\endgroup\@href {#1}{\urlprefix }}%
\providecommand \urlprefix  [0]{URL }%
\providecommand \Eprint [0]{\href }%
\providecommand \doibase [0]{http://dx.doi.org/}%
\providecommand \selectlanguage [0]{\@gobble}%
\providecommand \bibinfo  [0]{\@secondoftwo}%
\providecommand \bibfield  [0]{\@secondoftwo}%
\providecommand \translation [1]{[#1]}%
\providecommand \BibitemOpen [0]{}%
\providecommand \bibitemStop [0]{}%
\providecommand \bibitemNoStop [0]{.\EOS\space}%
\providecommand \EOS [0]{\spacefactor3000\relax}%
\providecommand \BibitemShut  [1]{\csname bibitem#1\endcsname}%
\let\auto@bib@innerbib\@empty
\bibitem [{\citenamefont {Bakker}\ and\ \citenamefont
  {Skinner}(2010)}]{Bakker2010}%
  \BibitemOpen
  \bibfield  {author} {\bibinfo {author} {\bibfnamefont {H.~J.}\ \bibnamefont
  {Bakker}}\ and\ \bibinfo {author} {\bibfnamefont {J.~L.}\ \bibnamefont
  {Skinner}},\ }\href {\doibase 10.1021/cr9001879} {\bibfield  {journal}
  {\bibinfo  {journal} {Chem. Rev.}\ }\textbf {\bibinfo {volume} {110}},\
  \bibinfo {pages} {1498} (\bibinfo {year} {2010})}\BibitemShut {NoStop}%
\bibitem [{\citenamefont {Scherer}\ \emph {et~al.}(1974)\citenamefont
  {Scherer}, \citenamefont {Go},\ and\ \citenamefont {Kint}}]{Scherer1974}%
  \BibitemOpen
  \bibfield  {author} {\bibinfo {author} {\bibfnamefont {J.~R.}\ \bibnamefont
  {Scherer}}, \bibinfo {author} {\bibfnamefont {M.~K.}\ \bibnamefont {Go}}, \
  and\ \bibinfo {author} {\bibfnamefont {S.}~\bibnamefont {Kint}},\ }\href
  {\doibase 10.1021/j100606a013} {\bibfield  {journal} {\bibinfo  {journal} {J.
  Phys. Chem.}\ }\textbf {\bibinfo {volume} {78}},\ \bibinfo {pages} {1304}
  (\bibinfo {year} {1974})}\BibitemShut {NoStop}%
\bibitem [{\citenamefont {Zhelyaskov}\ \emph {et~al.}(1989)\citenamefont
  {Zhelyaskov}, \citenamefont {Georgiev}, \citenamefont {Nickolov},\ and\
  \citenamefont {Miteva}}]{Zhelyaskov1989}%
  \BibitemOpen
  \bibfield  {author} {\bibinfo {author} {\bibfnamefont {V.}~\bibnamefont
  {Zhelyaskov}}, \bibinfo {author} {\bibfnamefont {G.}~\bibnamefont
  {Georgiev}}, \bibinfo {author} {\bibfnamefont {Z.}~\bibnamefont {Nickolov}},
  \ and\ \bibinfo {author} {\bibfnamefont {M.}~\bibnamefont {Miteva}},\ }\href
  {\doibase 10.1002/jrs.1250200203} {\bibfield  {journal} {\bibinfo  {journal}
  {J. Raman Spectrosc.}\ }\textbf {\bibinfo {volume} {20}},\ \bibinfo {pages}
  {67} (\bibinfo {year} {1989})}\BibitemShut {NoStop}%
\bibitem [{\citenamefont {Bray}\ \emph {et~al.}(2013)\citenamefont {Bray},
  \citenamefont {Chapman},\ and\ \citenamefont {Plakhotnik}}]{Bray2013}%
  \BibitemOpen
  \bibfield  {author} {\bibinfo {author} {\bibfnamefont {A.}~\bibnamefont
  {Bray}}, \bibinfo {author} {\bibfnamefont {R.}~\bibnamefont {Chapman}}, \
  and\ \bibinfo {author} {\bibfnamefont {T.}~\bibnamefont {Plakhotnik}},\
  }\href {\doibase 10.1364/AO.52.002503} {\bibfield  {journal} {\bibinfo
  {journal} {Appl. Opt.}\ }\textbf {\bibinfo {volume} {52}},\ \bibinfo {pages}
  {2503} (\bibinfo {year} {2013})}\BibitemShut {NoStop}%
\bibitem [{\citenamefont {De~Santis}\ \emph {et~al.}(1987)\citenamefont
  {De~Santis}, \citenamefont {Frattini}, \citenamefont {Sampoli}, \citenamefont
  {Mazzacurati}, \citenamefont {Nardone}, \citenamefont {Ricci},\ and\
  \citenamefont {Ruocco}}]{DeSantis1987}%
  \BibitemOpen
  \bibfield  {author} {\bibinfo {author} {\bibfnamefont {A.}~\bibnamefont
  {De~Santis}}, \bibinfo {author} {\bibfnamefont {R.}~\bibnamefont {Frattini}},
  \bibinfo {author} {\bibfnamefont {M.}~\bibnamefont {Sampoli}}, \bibinfo
  {author} {\bibfnamefont {V.}~\bibnamefont {Mazzacurati}}, \bibinfo {author}
  {\bibfnamefont {M.}~\bibnamefont {Nardone}}, \bibinfo {author} {\bibfnamefont
  {M.~A.}\ \bibnamefont {Ricci}}, \ and\ \bibinfo {author} {\bibfnamefont
  {G.}~\bibnamefont {Ruocco}},\ }\href {\doibase 10.1080/00268978700101741}
  {\bibfield  {journal} {\bibinfo  {journal} {Mol. Phys.}\ }\textbf {\bibinfo
  {volume} {61}},\ \bibinfo {pages} {1199} (\bibinfo {year}
  {1987})}\BibitemShut {NoStop}%
\bibitem [{\citenamefont {Walrafen}(1974)}]{Walrafen1973}%
  \BibitemOpen
  \bibfield  {author} {\bibinfo {author} {\bibfnamefont {G.~E.}\ \bibnamefont
  {Walrafen}},\ }in\ \href@noop {} {\emph {\bibinfo {booktitle} {Structure of
  Water and Aqueous Solutions}}},\ \bibinfo {editor} {edited by\ \bibinfo
  {editor} {\bibfnamefont {W.~A.~P.}\ \bibnamefont {Luck}}}\ (\bibinfo
  {publisher} {Verlag Chemie},\ \bibinfo {address} {Weinheim},\ \bibinfo {year}
  {1974})\ pp.\ \bibinfo {pages} {301--321}\BibitemShut {NoStop}%
\bibitem [{\citenamefont {Murphy}\ and\ \citenamefont
  {Bernstein}(1972)}]{Murphy1972}%
  \BibitemOpen
  \bibfield  {author} {\bibinfo {author} {\bibfnamefont {W.~F.}\ \bibnamefont
  {Murphy}}\ and\ \bibinfo {author} {\bibfnamefont {H.~J.}\ \bibnamefont
  {Bernstein}},\ }\href {\doibase 10.1021/j100652a010} {\bibfield  {journal}
  {\bibinfo  {journal} {J. Phys. Chem.}\ }\textbf {\bibinfo {volume} {76}},\
  \bibinfo {pages} {1147} (\bibinfo {year} {1972})}\BibitemShut {NoStop}%
\bibitem [{\citenamefont {Hare}\ and\ \citenamefont
  {Sorensen}(1990)}]{Hare1990}%
  \BibitemOpen
  \bibfield  {author} {\bibinfo {author} {\bibfnamefont {D.~E.}\ \bibnamefont
  {Hare}}\ and\ \bibinfo {author} {\bibfnamefont {C.~M.}\ \bibnamefont
  {Sorensen}},\ }\href {\doibase 10.1063/1.459599} {\bibfield  {journal}
  {\bibinfo  {journal} {J. Chem. Phys.}\ }\textbf {\bibinfo {volume} {93}},\
  \bibinfo {pages} {25} (\bibinfo {year} {1990})}\BibitemShut {NoStop}%
\bibitem [{\citenamefont {Walrafen}\ \emph
  {et~al.}(1986{\natexlab{a}})\citenamefont {Walrafen}, \citenamefont {Fisher},
  \citenamefont {Hokmabadi},\ and\ \citenamefont {Yang}}]{Walrafen1986}%
  \BibitemOpen
  \bibfield  {author} {\bibinfo {author} {\bibfnamefont {G.~E.}\ \bibnamefont
  {Walrafen}}, \bibinfo {author} {\bibfnamefont {M.~R.}\ \bibnamefont
  {Fisher}}, \bibinfo {author} {\bibfnamefont {M.~S.}\ \bibnamefont
  {Hokmabadi}}, \ and\ \bibinfo {author} {\bibfnamefont {W.}~\bibnamefont
  {Yang}},\ }\href {\doibase 10.1063/1.451384} {\bibfield  {journal} {\bibinfo
  {journal} {J. Chem. Phys.}\ }\textbf {\bibinfo {volume} {85}},\ \bibinfo
  {pages} {6970} (\bibinfo {year} {1986}{\natexlab{a}})}\BibitemShut {NoStop}%
\bibitem [{\citenamefont {Wan}\ \emph {et~al.}(2013)\citenamefont {Wan},
  \citenamefont {Spanu}, \citenamefont {Galli},\ and\ \citenamefont
  {Gygi}}]{Wan2013a}%
  \BibitemOpen
  \bibfield  {author} {\bibinfo {author} {\bibfnamefont {Q.}~\bibnamefont
  {Wan}}, \bibinfo {author} {\bibfnamefont {L.}~\bibnamefont {Spanu}}, \bibinfo
  {author} {\bibfnamefont {G.~A.}\ \bibnamefont {Galli}}, \ and\ \bibinfo
  {author} {\bibfnamefont {F.}~\bibnamefont {Gygi}},\ }\href {\doibase
  10.1021/ct4005307} {\bibfield  {journal} {\bibinfo  {journal} {J. Chem.
  Theory Comput.}\ }\textbf {\bibinfo {volume} {9}},\ \bibinfo {pages} {4124}
  (\bibinfo {year} {2013})}\BibitemShut {NoStop}%
\bibitem [{\citenamefont {Medders}\ and\ \citenamefont
  {Paesani}(2015)}]{Medders2015}%
  \BibitemOpen
  \bibfield  {author} {\bibinfo {author} {\bibfnamefont {G.~R.}\ \bibnamefont
  {Medders}}\ and\ \bibinfo {author} {\bibfnamefont {F.}~\bibnamefont
  {Paesani}},\ }\href {\doibase 10.1021/ct501131j} {\bibfield  {journal}
  {\bibinfo  {journal} {J. Chem. Theory Comput.}\ }\textbf {\bibinfo {volume}
  {11}},\ \bibinfo {pages} {1145} (\bibinfo {year} {2015})}\BibitemShut
  {NoStop}%
\bibitem [{\citenamefont {Marsalek}\ and\ \citenamefont
  {Markland}(2017)}]{Marsalek2017}%
  \BibitemOpen
  \bibfield  {author} {\bibinfo {author} {\bibfnamefont {O.}~\bibnamefont
  {Marsalek}}\ and\ \bibinfo {author} {\bibfnamefont {T.~E.}\ \bibnamefont
  {Markland}},\ }\href {\doibase 10.1021/acs.jpclett.7b00391} {\bibfield
  {journal} {\bibinfo  {journal} {J. Phys. Chem. Lett.}\ }\textbf {\bibinfo
  {volume} {8}},\ \bibinfo {pages} {1545} (\bibinfo {year} {2017})}\BibitemShut
  {NoStop}%
\bibitem [{\citenamefont {D’Arrigo}\ \emph {et~al.}(1981)\citenamefont
  {D’Arrigo}, \citenamefont {Maisano}, \citenamefont {Mallamace},
  \citenamefont {Migliardo},\ and\ \citenamefont {Wanderlingh}}]{DArrigo1981}%
  \BibitemOpen
  \bibfield  {author} {\bibinfo {author} {\bibfnamefont {G.}~\bibnamefont
  {D’Arrigo}}, \bibinfo {author} {\bibfnamefont {G.}~\bibnamefont {Maisano}},
  \bibinfo {author} {\bibfnamefont {F.}~\bibnamefont {Mallamace}}, \bibinfo
  {author} {\bibfnamefont {P.}~\bibnamefont {Migliardo}}, \ and\ \bibinfo
  {author} {\bibfnamefont {F.}~\bibnamefont {Wanderlingh}},\ }\href {\doibase
  10.1063/1.442629} {\bibfield  {journal} {\bibinfo  {journal} {J. Chem.
  Phys.}\ }\textbf {\bibinfo {volume} {75}},\ \bibinfo {pages} {4264} (\bibinfo
  {year} {1981})}\BibitemShut {NoStop}%
\bibitem [{\citenamefont {Walrafen}\ \emph
  {et~al.}(1986{\natexlab{b}})\citenamefont {Walrafen}, \citenamefont
  {Hokmabadi},\ and\ \citenamefont {Yang}}]{Walrafen1986a}%
  \BibitemOpen
  \bibfield  {author} {\bibinfo {author} {\bibfnamefont {G.~E.}\ \bibnamefont
  {Walrafen}}, \bibinfo {author} {\bibfnamefont {M.~S.}\ \bibnamefont
  {Hokmabadi}}, \ and\ \bibinfo {author} {\bibfnamefont {W.}~\bibnamefont
  {Yang}},\ }\href {\doibase 10.1063/1.451383} {\bibfield  {journal} {\bibinfo
  {journal} {J. Chem. Phys.}\ }\textbf {\bibinfo {volume} {85}},\ \bibinfo
  {pages} {6964} (\bibinfo {year} {1986}{\natexlab{b}})}\BibitemShut {NoStop}%
\bibitem [{\citenamefont {Sun}(2013)}]{Sun2013}%
  \BibitemOpen
  \bibfield  {author} {\bibinfo {author} {\bibfnamefont {Q.}~\bibnamefont
  {Sun}},\ }\href {\doibase 10.1016/j.cplett.2013.03.065} {\bibfield  {journal}
  {\bibinfo  {journal} {Chem. Phys. Lett.}\ }\textbf {\bibinfo {volume}
  {568-569}},\ \bibinfo {pages} {90} (\bibinfo {year} {2013})}\BibitemShut
  {NoStop}%
\bibitem [{\citenamefont {Harada}\ \emph {et~al.}(2017)\citenamefont {Harada},
  \citenamefont {Miyawaki}, \citenamefont {Niwa}, \citenamefont {Yamazoe},
  \citenamefont {Pettersson},\ and\ \citenamefont {Nilsson}}]{Harada2017}%
  \BibitemOpen
  \bibfield  {author} {\bibinfo {author} {\bibfnamefont {Y.}~\bibnamefont
  {Harada}}, \bibinfo {author} {\bibfnamefont {J.}~\bibnamefont {Miyawaki}},
  \bibinfo {author} {\bibfnamefont {H.}~\bibnamefont {Niwa}}, \bibinfo {author}
  {\bibfnamefont {K.}~\bibnamefont {Yamazoe}}, \bibinfo {author} {\bibfnamefont
  {L.~G.~M.}\ \bibnamefont {Pettersson}}, \ and\ \bibinfo {author}
  {\bibfnamefont {A.}~\bibnamefont {Nilsson}},\ }\href {\doibase
  10.1021/acs.jpclett.7b02060} {\bibfield  {journal} {\bibinfo  {journal} {J.
  Phys. Chem. Lett.}\ }\textbf {\bibinfo {volume} {8}},\ \bibinfo {pages}
  {5487} (\bibinfo {year} {2017})}\BibitemShut {NoStop}%
\bibitem [{\citenamefont {Geissler}(2005)}]{Geissler2005}%
  \BibitemOpen
  \bibfield  {author} {\bibinfo {author} {\bibfnamefont {P.~L.}\ \bibnamefont
  {Geissler}},\ }\href {\doibase 10.1021/ja0545214} {\bibfield  {journal}
  {\bibinfo  {journal} {J. Am. Chem. Soc.}\ }\textbf {\bibinfo {volume}
  {127}},\ \bibinfo {pages} {14930} (\bibinfo {year} {2005})}\BibitemShut
  {NoStop}%
\bibitem [{\citenamefont {Smith}\ \emph {et~al.}(2005)\citenamefont {Smith},
  \citenamefont {Cappa}, \citenamefont {Wilson}, \citenamefont {Cohen},
  \citenamefont {Geissler},\ and\ \citenamefont {Saykally}}]{Smith2005}%
  \BibitemOpen
  \bibfield  {author} {\bibinfo {author} {\bibfnamefont {J.~D.}\ \bibnamefont
  {Smith}}, \bibinfo {author} {\bibfnamefont {C.~D.}\ \bibnamefont {Cappa}},
  \bibinfo {author} {\bibfnamefont {K.~R.}\ \bibnamefont {Wilson}}, \bibinfo
  {author} {\bibfnamefont {R.~C.}\ \bibnamefont {Cohen}}, \bibinfo {author}
  {\bibfnamefont {P.~L.}\ \bibnamefont {Geissler}}, \ and\ \bibinfo {author}
  {\bibfnamefont {R.~J.}\ \bibnamefont {Saykally}},\ }\href {\doibase
  10.1073/pnas.0506899102} {\bibfield  {journal} {\bibinfo  {journal} {Proc.
  Natl. Acad. Sci. U.S.A.}\ }\textbf {\bibinfo {volume} {102}},\ \bibinfo
  {pages} {14171} (\bibinfo {year} {2005})}\BibitemShut {NoStop}%
\bibitem [{\citenamefont {Corcelli}\ \emph {et~al.}(2004)\citenamefont
  {Corcelli}, \citenamefont {Lawrence},\ and\ \citenamefont
  {Skinner}}]{Corcelli2004}%
  \BibitemOpen
  \bibfield  {author} {\bibinfo {author} {\bibfnamefont {S.~A.}\ \bibnamefont
  {Corcelli}}, \bibinfo {author} {\bibfnamefont {C.~P.}\ \bibnamefont
  {Lawrence}}, \ and\ \bibinfo {author} {\bibfnamefont {J.~L.}\ \bibnamefont
  {Skinner}},\ }\href {\doibase 10.1063/1.1683072} {\bibfield  {journal}
  {\bibinfo  {journal} {J. Chem. Phys.}\ }\textbf {\bibinfo {volume} {120}},\
  \bibinfo {pages} {8107} (\bibinfo {year} {2004})}\BibitemShut {NoStop}%
\bibitem [{\citenamefont {Corcelli}\ and\ \citenamefont
  {Skinner}(2005)}]{Corcelli2005}%
  \BibitemOpen
  \bibfield  {author} {\bibinfo {author} {\bibfnamefont {S.~A.}\ \bibnamefont
  {Corcelli}}\ and\ \bibinfo {author} {\bibfnamefont {J.~L.}\ \bibnamefont
  {Skinner}},\ }\href {\doibase 10.1021/jp0506540} {\bibfield  {journal}
  {\bibinfo  {journal} {J. Phys. Chem. A}\ }\textbf {\bibinfo {volume} {109}},\
  \bibinfo {pages} {6154} (\bibinfo {year} {2005})}\BibitemShut {NoStop}%
\bibitem [{\citenamefont {Auer}\ and\ \citenamefont
  {Skinner}(2008)}]{Auer2008}%
  \BibitemOpen
  \bibfield  {author} {\bibinfo {author} {\bibfnamefont {B.~M.}\ \bibnamefont
  {Auer}}\ and\ \bibinfo {author} {\bibfnamefont {J.~L.}\ \bibnamefont
  {Skinner}},\ }\href {\doibase 10.1063/1.2925258} {\bibfield  {journal}
  {\bibinfo  {journal} {J. Chem. Phys.}\ }\textbf {\bibinfo {volume} {128}},\
  \bibinfo {pages} {224511} (\bibinfo {year} {2008})}\BibitemShut {NoStop}%
\bibitem [{\citenamefont {Yang}\ and\ \citenamefont
  {Skinner}(2010)}]{Yang2010}%
  \BibitemOpen
  \bibfield  {author} {\bibinfo {author} {\bibfnamefont {M.}~\bibnamefont
  {Yang}}\ and\ \bibinfo {author} {\bibfnamefont {J.~L.}\ \bibnamefont
  {Skinner}},\ }\href {\doibase 10.1039/b918314k} {\bibfield  {journal}
  {\bibinfo  {journal} {Phys. Chem. Chem. Phys.}\ }\textbf {\bibinfo {volume}
  {12}},\ \bibinfo {pages} {982} (\bibinfo {year} {2010})}\BibitemShut
  {NoStop}%
\bibitem [{\citenamefont {Tainter}\ \emph {et~al.}(2013)\citenamefont
  {Tainter}, \citenamefont {Ni}, \citenamefont {Shi},\ and\ \citenamefont
  {Skinner}}]{Tainter2013}%
  \BibitemOpen
  \bibfield  {author} {\bibinfo {author} {\bibfnamefont {C.~J.}\ \bibnamefont
  {Tainter}}, \bibinfo {author} {\bibfnamefont {Y.}~\bibnamefont {Ni}},
  \bibinfo {author} {\bibfnamefont {L.}~\bibnamefont {Shi}}, \ and\ \bibinfo
  {author} {\bibfnamefont {J.~L.}\ \bibnamefont {Skinner}},\ }\href {\doibase
  10.1021/jz301780k} {\bibfield  {journal} {\bibinfo  {journal} {J. Phys. Chem.
  Lett.}\ }\textbf {\bibinfo {volume} {4}},\ \bibinfo {pages} {12} (\bibinfo
  {year} {2013})}\BibitemShut {NoStop}%
\bibitem [{\citenamefont {Madden}\ and\ \citenamefont
  {Impey}(1986)}]{Madden1986}%
  \BibitemOpen
  \bibfield  {author} {\bibinfo {author} {\bibfnamefont {P.~A.}\ \bibnamefont
  {Madden}}\ and\ \bibinfo {author} {\bibfnamefont {R.~W.}\ \bibnamefont
  {Impey}},\ }\href {\doibase 10.1016/0009-2614(86)80051-3} {\bibfield
  {journal} {\bibinfo  {journal} {Chem. Phys. Lett.}\ }\textbf {\bibinfo
  {volume} {123}},\ \bibinfo {pages} {502} (\bibinfo {year}
  {1986})}\BibitemShut {NoStop}%
\bibitem [{\citenamefont {Mazzacurati}\ \emph {et~al.}(1989)\citenamefont
  {Mazzacurati}, \citenamefont {Ricci}, \citenamefont {Ruocco},\ and\
  \citenamefont {Sampoli}}]{Mazzacurati1989}%
  \BibitemOpen
  \bibfield  {author} {\bibinfo {author} {\bibfnamefont {V.}~\bibnamefont
  {Mazzacurati}}, \bibinfo {author} {\bibfnamefont {M.}~\bibnamefont {Ricci}},
  \bibinfo {author} {\bibfnamefont {G.}~\bibnamefont {Ruocco}}, \ and\ \bibinfo
  {author} {\bibfnamefont {M.}~\bibnamefont {Sampoli}},\ }\href {\doibase
  10.1016/0009-2614(89)87504-9} {\bibfield  {journal} {\bibinfo  {journal}
  {Chem. Phys. Lett.}\ }\textbf {\bibinfo {volume} {159}},\ \bibinfo {pages}
  {383} (\bibinfo {year} {1989})}\BibitemShut {NoStop}%
\bibitem [{\citenamefont {Bursulaya}\ and\ \citenamefont
  {Kim}(1998)}]{Bursulaya1998}%
  \BibitemOpen
  \bibfield  {author} {\bibinfo {author} {\bibfnamefont {B.~D.}\ \bibnamefont
  {Bursulaya}}\ and\ \bibinfo {author} {\bibfnamefont {H.~J.}\ \bibnamefont
  {Kim}},\ }\href {\doibase 10.1063/1.477102} {\bibfield  {journal} {\bibinfo
  {journal} {J. Chem. Phys.}\ }\textbf {\bibinfo {volume} {109}},\ \bibinfo
  {pages} {4911} (\bibinfo {year} {1998})}\BibitemShut {NoStop}%
\bibitem [{\citenamefont {Behler}\ and\ \citenamefont
  {Parrinello}(2007)}]{Behler2007}%
  \BibitemOpen
  \bibfield  {author} {\bibinfo {author} {\bibfnamefont {J.}~\bibnamefont
  {Behler}}\ and\ \bibinfo {author} {\bibfnamefont {M.}~\bibnamefont
  {Parrinello}},\ }\href {\doibase 10.1103/PhysRevLett.98.146401} {\bibfield
  {journal} {\bibinfo  {journal} {Phys. Rev. Lett.}\ }\textbf {\bibinfo
  {volume} {98}},\ \bibinfo {pages} {146401} (\bibinfo {year}
  {2007})}\BibitemShut {NoStop}%
\bibitem [{\citenamefont {Artrith}\ and\ \citenamefont
  {Urban}(2016)}]{Artrith2016}%
  \BibitemOpen
  \bibfield  {author} {\bibinfo {author} {\bibfnamefont {N.}~\bibnamefont
  {Artrith}}\ and\ \bibinfo {author} {\bibfnamefont {A.}~\bibnamefont
  {Urban}},\ }\href {\doibase 10.1016/j.commatsci.2015.11.047} {\bibfield
  {journal} {\bibinfo  {journal} {Comput. Mater. Sci.}\ }\textbf {\bibinfo
  {volume} {114}},\ \bibinfo {pages} {135} (\bibinfo {year}
  {2016})}\BibitemShut {NoStop}%
\bibitem [{\citenamefont {Behler}(2016)}]{Behler2016}%
  \BibitemOpen
  \bibfield  {author} {\bibinfo {author} {\bibfnamefont {J.}~\bibnamefont
  {Behler}},\ }\href {\doibase 10.1063/1.4966192} {\bibfield  {journal}
  {\bibinfo  {journal} {J. Chem. Phys.}\ }\textbf {\bibinfo {volume} {145}},\
  \bibinfo {pages} {170901} (\bibinfo {year} {2016})}\BibitemShut {NoStop}%
\bibitem [{\citenamefont {Morawietz}\ \emph {et~al.}(2016)\citenamefont
  {Morawietz}, \citenamefont {Singraber}, \citenamefont {Dellago},\ and\
  \citenamefont {Behler}}]{Morawietz2016}%
  \BibitemOpen
  \bibfield  {author} {\bibinfo {author} {\bibfnamefont {T.}~\bibnamefont
  {Morawietz}}, \bibinfo {author} {\bibfnamefont {A.}~\bibnamefont
  {Singraber}}, \bibinfo {author} {\bibfnamefont {C.}~\bibnamefont {Dellago}},
  \ and\ \bibinfo {author} {\bibfnamefont {J.}~\bibnamefont {Behler}},\ }\href
  {\doibase 10.1073/pnas.1602375113} {\bibfield  {journal} {\bibinfo  {journal}
  {Proc. Natl. Acad. Sci. U.S.A.}\ }\textbf {\bibinfo {volume} {113}},\
  \bibinfo {pages} {8368} (\bibinfo {year} {2016})}\BibitemShut {NoStop}%
\bibitem [{\citenamefont {Heyden}\ \emph {et~al.}(2008)\citenamefont {Heyden},
  \citenamefont {Br{\"{u}}ndermann}, \citenamefont {Heugen}, \citenamefont
  {Niehues}, \citenamefont {Leitner},\ and\ \citenamefont
  {Havenith}}]{Heyden2008}%
  \BibitemOpen
  \bibfield  {author} {\bibinfo {author} {\bibfnamefont {M.}~\bibnamefont
  {Heyden}}, \bibinfo {author} {\bibfnamefont {E.}~\bibnamefont
  {Br{\"{u}}ndermann}}, \bibinfo {author} {\bibfnamefont {U.}~\bibnamefont
  {Heugen}}, \bibinfo {author} {\bibfnamefont {G.}~\bibnamefont {Niehues}},
  \bibinfo {author} {\bibfnamefont {D.~M.}\ \bibnamefont {Leitner}}, \ and\
  \bibinfo {author} {\bibfnamefont {M.}~\bibnamefont {Havenith}},\ }\href
  {\doibase 10.1021/ja00781083} {\bibfield  {journal} {\bibinfo  {journal} {J.
  Am. Chem. Soc.}\ }\textbf {\bibinfo {volume} {130}},\ \bibinfo {pages} {5773}
  (\bibinfo {year} {2008})}\BibitemShut {NoStop}%
\bibitem [{\citenamefont {Artrith}\ and\ \citenamefont
  {Kolpak}(2014)}]{Artrith2014}%
  \BibitemOpen
  \bibfield  {author} {\bibinfo {author} {\bibfnamefont {N.}~\bibnamefont
  {Artrith}}\ and\ \bibinfo {author} {\bibfnamefont {A.~M.}\ \bibnamefont
  {Kolpak}},\ }\href {\doibase 10.1021/nl5005674} {\bibfield  {journal}
  {\bibinfo  {journal} {Nano Lett.}\ }\textbf {\bibinfo {volume} {14}},\
  \bibinfo {pages} {2670} (\bibinfo {year} {2014})}\BibitemShut {NoStop}%
\bibitem [{\citenamefont {Gierszal}\ \emph {et~al.}(2011)\citenamefont
  {Gierszal}, \citenamefont {Davis}, \citenamefont {Hands}, \citenamefont
  {Wilcox}, \citenamefont {Slipchenko},\ and\ \citenamefont
  {Ben-Amotz}}]{Gierszal2011}%
  \BibitemOpen
  \bibfield  {author} {\bibinfo {author} {\bibfnamefont {K.~P.}\ \bibnamefont
  {Gierszal}}, \bibinfo {author} {\bibfnamefont {J.~G.}\ \bibnamefont {Davis}},
  \bibinfo {author} {\bibfnamefont {M.~D.}\ \bibnamefont {Hands}}, \bibinfo
  {author} {\bibfnamefont {D.~S.}\ \bibnamefont {Wilcox}}, \bibinfo {author}
  {\bibfnamefont {L.~V.}\ \bibnamefont {Slipchenko}}, \ and\ \bibinfo {author}
  {\bibfnamefont {D.}~\bibnamefont {Ben-Amotz}},\ }\href {\doibase
  10.1021/jz201373e} {\bibfield  {journal} {\bibinfo  {journal} {J. Phys. Chem.
  Lett.}\ }\textbf {\bibinfo {volume} {2}},\ \bibinfo {pages} {2930} (\bibinfo
  {year} {2011})}\BibitemShut {NoStop}%
\bibitem [{\citenamefont {Fayer}\ and\ \citenamefont
  {Levinger}(2010)}]{Fayer2010}%
  \BibitemOpen
  \bibfield  {author} {\bibinfo {author} {\bibfnamefont {M.~D.}\ \bibnamefont
  {Fayer}}\ and\ \bibinfo {author} {\bibfnamefont {N.~E.}\ \bibnamefont
  {Levinger}},\ }\href {\doibase 10.1146/annurev-anchem-070109-103410}
  {\bibfield  {journal} {\bibinfo  {journal} {Annu. Rev. Anal. Chem.}\ }\textbf
  {\bibinfo {volume} {3}},\ \bibinfo {pages} {89} (\bibinfo {year}
  {2010})}\BibitemShut {NoStop}%
\bibitem [{\citenamefont {Crans}\ and\ \citenamefont
  {Levinger}(2012)}]{Crans2012}%
  \BibitemOpen
  \bibfield  {author} {\bibinfo {author} {\bibfnamefont {D.~C.}\ \bibnamefont
  {Crans}}\ and\ \bibinfo {author} {\bibfnamefont {N.~E.}\ \bibnamefont
  {Levinger}},\ }\href {\doibase 10.1021/ar200269g} {\bibfield  {journal}
  {\bibinfo  {journal} {Acc. Chem. Res.}\ }\textbf {\bibinfo {volume} {45}},\
  \bibinfo {pages} {1637} (\bibinfo {year} {2012})}\BibitemShut {NoStop}%
\bibitem [{\citenamefont {Davis}\ \emph {et~al.}(2012)\citenamefont {Davis},
  \citenamefont {Gierszal}, \citenamefont {Wang},\ and\ \citenamefont
  {Ben-Amotz}}]{Davis2012a}%
  \BibitemOpen
  \bibfield  {author} {\bibinfo {author} {\bibfnamefont {J.~G.}\ \bibnamefont
  {Davis}}, \bibinfo {author} {\bibfnamefont {K.~P.}\ \bibnamefont {Gierszal}},
  \bibinfo {author} {\bibfnamefont {P.}~\bibnamefont {Wang}}, \ and\ \bibinfo
  {author} {\bibfnamefont {D.}~\bibnamefont {Ben-Amotz}},\ }\href {\doibase
  10.1038/nature11570} {\bibfield  {journal} {\bibinfo  {journal} {Nature}\
  }\textbf {\bibinfo {volume} {491}},\ \bibinfo {pages} {582} (\bibinfo {year}
  {2012})}\BibitemShut {NoStop}%
\bibitem [{\citenamefont {Bakulin}\ \emph {et~al.}(2013)\citenamefont
  {Bakulin}, \citenamefont {Cringus}, \citenamefont {Pieniazek}, \citenamefont
  {Skinner}, \citenamefont {Jansen},\ and\ \citenamefont
  {Pshenichnikov}}]{Bakulin2013}%
  \BibitemOpen
  \bibfield  {author} {\bibinfo {author} {\bibfnamefont {A.~A.}\ \bibnamefont
  {Bakulin}}, \bibinfo {author} {\bibfnamefont {D.}~\bibnamefont {Cringus}},
  \bibinfo {author} {\bibfnamefont {P.~A.}\ \bibnamefont {Pieniazek}}, \bibinfo
  {author} {\bibfnamefont {J.~L.}\ \bibnamefont {Skinner}}, \bibinfo {author}
  {\bibfnamefont {T.~L.~C.}\ \bibnamefont {Jansen}}, \ and\ \bibinfo {author}
  {\bibfnamefont {M.~S.}\ \bibnamefont {Pshenichnikov}},\ }\href {\doibase
  10.1021/jp405853j} {\bibfield  {journal} {\bibinfo  {journal} {J. Phys. Chem.
  B}\ }\textbf {\bibinfo {volume} {117}},\ \bibinfo {pages} {15545} (\bibinfo
  {year} {2013})}\BibitemShut {NoStop}%
\bibitem [{\citenamefont {Russo}\ \emph {et~al.}(2017)\citenamefont {Russo},
  \citenamefont {Laloni}, \citenamefont {Filabozzi},\ and\ \citenamefont
  {Heyden}}]{Russo2017}%
  \BibitemOpen
  \bibfield  {author} {\bibinfo {author} {\bibfnamefont {D.}~\bibnamefont
  {Russo}}, \bibinfo {author} {\bibfnamefont {A.}~\bibnamefont {Laloni}},
  \bibinfo {author} {\bibfnamefont {A.}~\bibnamefont {Filabozzi}}, \ and\
  \bibinfo {author} {\bibfnamefont {M.}~\bibnamefont {Heyden}},\ }\href
  {\doibase 10.1073/pnas.1705279114} {\bibfield  {journal} {\bibinfo  {journal}
  {Proc. Natl. Acad. Sci. U.S.A.}\ }\textbf {\bibinfo {volume} {114}},\
  \bibinfo {pages} {11410} (\bibinfo {year} {2017})}\BibitemShut {NoStop}%
\bibitem [{\citenamefont {Brooker}\ \emph {et~al.}(1988)\citenamefont
  {Brooker}, \citenamefont {Nielsen},\ and\ \citenamefont
  {Praestgaard}}]{Brooker1988}%
  \BibitemOpen
  \bibfield  {author} {\bibinfo {author} {\bibfnamefont {M.~H.}\ \bibnamefont
  {Brooker}}, \bibinfo {author} {\bibfnamefont {O.~F.}\ \bibnamefont
  {Nielsen}}, \ and\ \bibinfo {author} {\bibfnamefont {E.}~\bibnamefont
  {Praestgaard}},\ }\href {\doibase 10.1002/jrs.1250190202} {\bibfield
  {journal} {\bibinfo  {journal} {J. Raman Spectrosc.}\ }\textbf {\bibinfo
  {volume} {19}},\ \bibinfo {pages} {71} (\bibinfo {year} {1988})}\BibitemShut
  {NoStop}%
\bibitem [{\citenamefont {Brooker}\ \emph {et~al.}(1989)\citenamefont
  {Brooker}, \citenamefont {Hancock}, \citenamefont {Rice},\ and\ \citenamefont
  {Shapter}}]{Brooker1989}%
  \BibitemOpen
  \bibfield  {author} {\bibinfo {author} {\bibfnamefont {M.~H.}\ \bibnamefont
  {Brooker}}, \bibinfo {author} {\bibfnamefont {G.}~\bibnamefont {Hancock}},
  \bibinfo {author} {\bibfnamefont {B.~C.}\ \bibnamefont {Rice}}, \ and\
  \bibinfo {author} {\bibfnamefont {J.}~\bibnamefont {Shapter}},\ }\href
  {\doibase 10.1002/jrs.1250201009} {\bibfield  {journal} {\bibinfo  {journal}
  {J. Raman Spectrosc.}\ }\textbf {\bibinfo {volume} {20}},\ \bibinfo {pages}
  {683} (\bibinfo {year} {1989})}\BibitemShut {NoStop}%
\bibitem [{\citenamefont {Mar{\'{e}}chal}(2011)}]{Marechal2011}%
  \BibitemOpen
  \bibfield  {author} {\bibinfo {author} {\bibfnamefont {Y.}~\bibnamefont
  {Mar{\'{e}}chal}},\ }\href {\doibase 10.1016/j.molstruc.2011.07.054}
  {\bibfield  {journal} {\bibinfo  {journal} {J. Mol. Struct.}\ }\textbf
  {\bibinfo {volume} {1004}},\ \bibinfo {pages} {146} (\bibinfo {year}
  {2011})}\BibitemShut {NoStop}%
\bibitem [{\citenamefont {Noda}(1993)}]{Noda1993}%
  \BibitemOpen
  \bibfield  {author} {\bibinfo {author} {\bibfnamefont {I.}~\bibnamefont
  {Noda}},\ }\href {\doibase 10.1366/0003702934067694} {\bibfield  {journal}
  {\bibinfo  {journal} {Appl. Spectrosc.}\ }\textbf {\bibinfo {volume} {47}},\
  \bibinfo {pages} {1329} (\bibinfo {year} {1993})}\BibitemShut {NoStop}%
\bibitem [{\citenamefont {Walrafen}\ and\ \citenamefont
  {Pugh}(2004)}]{Walrafen2004}%
  \BibitemOpen
  \bibfield  {author} {\bibinfo {author} {\bibfnamefont {G.~E.}\ \bibnamefont
  {Walrafen}}\ and\ \bibinfo {author} {\bibfnamefont {E.}~\bibnamefont
  {Pugh}},\ }\href {\doibase 10.1023/B:JOSL.0000026646.33891.a8} {\bibfield
  {journal} {\bibinfo  {journal} {J. Solution Chem.}\ }\textbf {\bibinfo
  {volume} {33}},\ \bibinfo {pages} {81} (\bibinfo {year} {2004})}\BibitemShut
  {NoStop}%
\bibitem [{\citenamefont {Mar{\'{e}}chal}(1991)}]{Marechal1991}%
  \BibitemOpen
  \bibfield  {author} {\bibinfo {author} {\bibfnamefont {Y.}~\bibnamefont
  {Mar{\'{e}}chal}},\ }\href {\doibase 10.1063/1.461630} {\bibfield  {journal}
  {\bibinfo  {journal} {J. Chem. Phys.}\ }\textbf {\bibinfo {volume} {95}},\
  \bibinfo {pages} {5565} (\bibinfo {year} {1991})}\BibitemShut {NoStop}%
\bibitem [{\citenamefont {McCoy}(2014)}]{McCoy2014}%
  \BibitemOpen
  \bibfield  {author} {\bibinfo {author} {\bibfnamefont {A.~B.}\ \bibnamefont
  {McCoy}},\ }\href {\doibase 10.1021/jp501647e} {\bibfield  {journal}
  {\bibinfo  {journal} {J. Phys. Chem. B}\ }\textbf {\bibinfo {volume} {118}},\
  \bibinfo {pages} {8286} (\bibinfo {year} {2014})}\BibitemShut {NoStop}%
\bibitem [{\citenamefont {Devlin}\ \emph {et~al.}(2001)\citenamefont {Devlin},
  \citenamefont {Sadlej},\ and\ \citenamefont {Buch}}]{Devlin2001}%
  \BibitemOpen
  \bibfield  {author} {\bibinfo {author} {\bibfnamefont {J.~P.}\ \bibnamefont
  {Devlin}}, \bibinfo {author} {\bibfnamefont {J.}~\bibnamefont {Sadlej}}, \
  and\ \bibinfo {author} {\bibfnamefont {V.}~\bibnamefont {Buch}},\ }\href
  {\doibase 10.1021/jp003455j} {\bibfield  {journal} {\bibinfo  {journal} {J.
  Phys. Chem. A}\ }\textbf {\bibinfo {volume} {105}},\ \bibinfo {pages} {974}
  (\bibinfo {year} {2001})}\BibitemShut {NoStop}%
\bibitem [{\citenamefont {Lawton}\ and\ \citenamefont
  {Sylvestre}(1971)}]{Lawton1971}%
  \BibitemOpen
  \bibfield  {author} {\bibinfo {author} {\bibfnamefont {W.~H.}\ \bibnamefont
  {Lawton}}\ and\ \bibinfo {author} {\bibfnamefont {E.~A.}\ \bibnamefont
  {Sylvestre}},\ }\href {\doibase 10.1080/00401706.1971.10488823} {\bibfield
  {journal} {\bibinfo  {journal} {Technometrics}\ }\textbf {\bibinfo {volume}
  {13}},\ \bibinfo {pages} {617} (\bibinfo {year} {1971})}\BibitemShut
  {NoStop}%
\bibitem [{\citenamefont {Tauler}\ \emph {et~al.}(1995)\citenamefont {Tauler},
  \citenamefont {Smilde},\ and\ \citenamefont {Kowalski}}]{Tauler1995}%
  \BibitemOpen
  \bibfield  {author} {\bibinfo {author} {\bibfnamefont {R.}~\bibnamefont
  {Tauler}}, \bibinfo {author} {\bibfnamefont {A.}~\bibnamefont {Smilde}}, \
  and\ \bibinfo {author} {\bibfnamefont {B.}~\bibnamefont {Kowalski}},\ }\href
  {\doibase 10.1002/cem.1180090105} {\bibfield  {journal} {\bibinfo  {journal}
  {J. Chemom.}\ }\textbf {\bibinfo {volume} {9}},\ \bibinfo {pages} {31}
  (\bibinfo {year} {1995})}\BibitemShut {NoStop}%
\bibitem [{\citenamefont {Jiang}\ \emph {et~al.}(2004)\citenamefont {Jiang},
  \citenamefont {Liang},\ and\ \citenamefont {Ozaki}}]{Jiang2004}%
  \BibitemOpen
  \bibfield  {author} {\bibinfo {author} {\bibfnamefont {J.~H.}\ \bibnamefont
  {Jiang}}, \bibinfo {author} {\bibfnamefont {Y.}~\bibnamefont {Liang}}, \ and\
  \bibinfo {author} {\bibfnamefont {Y.}~\bibnamefont {Ozaki}},\ }\href
  {\doibase 10.1016/j.chemolab.2003.07.002} {\bibfield  {journal} {\bibinfo
  {journal} {Chemom. Intell. Lab. Syst.}\ }\textbf {\bibinfo {volume} {71}},\
  \bibinfo {pages} {1} (\bibinfo {year} {2004})}\BibitemShut {NoStop}%
\bibitem [{\citenamefont {Daly~Jr.}\ \emph {et~al.}(2017)\citenamefont
  {Daly~Jr.}, \citenamefont {Streacker}, \citenamefont {Sun}, \citenamefont
  {Pattenaude}, \citenamefont {Petersen}, \citenamefont {Corcelli},\ and\
  \citenamefont {Ben-Amotz}}]{Daly2017}%
  \BibitemOpen
  \bibfield  {author} {\bibinfo {author} {\bibfnamefont {C.~A.}\ \bibnamefont
  {Daly~Jr.}}, \bibinfo {author} {\bibfnamefont {L.~M.}\ \bibnamefont
  {Streacker}}, \bibinfo {author} {\bibfnamefont {Y.}~\bibnamefont {Sun}},
  \bibinfo {author} {\bibfnamefont {S.~R.}\ \bibnamefont {Pattenaude}},
  \bibinfo {author} {\bibfnamefont {P.~B.}\ \bibnamefont {Petersen}}, \bibinfo
  {author} {\bibfnamefont {S.~A.}\ \bibnamefont {Corcelli}}, \ and\ \bibinfo
  {author} {\bibfnamefont {D.}~\bibnamefont {Ben-Amotz}},\ }\href {\doibase
  10.1021/acs.jpclett.7b02435} {\bibfield  {journal} {\bibinfo  {journal} {J.
  Phys. Chem. Lett.}\ }\textbf {\bibinfo {volume} {8}},\ \bibinfo {pages}
  {5246} (\bibinfo {year} {2017})}\BibitemShut {NoStop}%
\bibitem [{\citenamefont {Rankin}\ and\ \citenamefont
  {Ben-Amotz}(2013)}]{Rankin2013}%
  \BibitemOpen
  \bibfield  {author} {\bibinfo {author} {\bibfnamefont {B.~M.}\ \bibnamefont
  {Rankin}}\ and\ \bibinfo {author} {\bibfnamefont {D.}~\bibnamefont
  {Ben-Amotz}},\ }\href {\doibase 10.1021/ja4036303} {\bibfield  {journal}
  {\bibinfo  {journal} {J. Am. Chem. Soc.}\ }\textbf {\bibinfo {volume}
  {135}},\ \bibinfo {pages} {8818} (\bibinfo {year} {2013})}\BibitemShut
  {NoStop}%
\bibitem [{\citenamefont {Zukowski}\ \emph {et~al.}(2017)\citenamefont
  {Zukowski}, \citenamefont {Mitev}, \citenamefont {Hermansson},\ and\
  \citenamefont {Ben-Amotz}}]{Zukowski2017}%
  \BibitemOpen
  \bibfield  {author} {\bibinfo {author} {\bibfnamefont {S.~R.}\ \bibnamefont
  {Zukowski}}, \bibinfo {author} {\bibfnamefont {P.~D.}\ \bibnamefont {Mitev}},
  \bibinfo {author} {\bibfnamefont {K.}~\bibnamefont {Hermansson}}, \ and\
  \bibinfo {author} {\bibfnamefont {D.}~\bibnamefont {Ben-Amotz}},\ }\href
  {\doibase 10.1021/acs.jpclett.7b00971} {\bibfield  {journal} {\bibinfo
  {journal} {J. Phys. Chem. Lett.}\ }\textbf {\bibinfo {volume} {8}},\ \bibinfo
  {pages} {2971} (\bibinfo {year} {2017})}\BibitemShut {NoStop}%
\bibitem [{\citenamefont {Rankin}\ \emph {et~al.}(2015)\citenamefont {Rankin},
  \citenamefont {Ben-Amotz}, \citenamefont {Van Der~Post},\ and\ \citenamefont
  {Bakker}}]{Rankin2015}%
  \BibitemOpen
  \bibfield  {author} {\bibinfo {author} {\bibfnamefont {B.~M.}\ \bibnamefont
  {Rankin}}, \bibinfo {author} {\bibfnamefont {D.}~\bibnamefont {Ben-Amotz}},
  \bibinfo {author} {\bibfnamefont {S.~T.}\ \bibnamefont {Van Der~Post}}, \
  and\ \bibinfo {author} {\bibfnamefont {H.~J.}\ \bibnamefont {Bakker}},\
  }\href {\doibase 10.1021/jz5027129} {\bibfield  {journal} {\bibinfo
  {journal} {J. Phys. Chem. Lett.}\ }\textbf {\bibinfo {volume} {6}},\ \bibinfo
  {pages} {688} (\bibinfo {year} {2015})}\BibitemShut {NoStop}%
\bibitem [{\citenamefont {Mochizuki}\ \emph {et~al.}(2016)\citenamefont
  {Mochizuki}, \citenamefont {Pattenaude},\ and\ \citenamefont
  {Ben-Amotz}}]{Mochizuki2016}%
  \BibitemOpen
  \bibfield  {author} {\bibinfo {author} {\bibfnamefont {K.}~\bibnamefont
  {Mochizuki}}, \bibinfo {author} {\bibfnamefont {S.~R.}\ \bibnamefont
  {Pattenaude}}, \ and\ \bibinfo {author} {\bibfnamefont {D.}~\bibnamefont
  {Ben-Amotz}},\ }\href {\doibase 10.1021/jacs.6b04914} {\bibfield  {journal}
  {\bibinfo  {journal} {J. Am. Chem. Soc.}\ }\textbf {\bibinfo {volume}
  {138}},\ \bibinfo {pages} {9045} (\bibinfo {year} {2016})}\BibitemShut
  {NoStop}%
\bibitem [{\citenamefont {Scheu}\ \emph {et~al.}(2014)\citenamefont {Scheu},
  \citenamefont {Rankin}, \citenamefont {Chen}, \citenamefont {Jena},
  \citenamefont {Ben-Amotz},\ and\ \citenamefont {Roke}}]{Scheu2014}%
  \BibitemOpen
  \bibfield  {author} {\bibinfo {author} {\bibfnamefont {R.}~\bibnamefont
  {Scheu}}, \bibinfo {author} {\bibfnamefont {B.~M.}\ \bibnamefont {Rankin}},
  \bibinfo {author} {\bibfnamefont {Y.}~\bibnamefont {Chen}}, \bibinfo {author}
  {\bibfnamefont {K.~C.}\ \bibnamefont {Jena}}, \bibinfo {author}
  {\bibfnamefont {D.}~\bibnamefont {Ben-Amotz}}, \ and\ \bibinfo {author}
  {\bibfnamefont {S.}~\bibnamefont {Roke}},\ }\href {\doibase
  10.1002/anie.201310266} {\bibfield  {journal} {\bibinfo  {journal} {Angew.
  Chem.}\ }\textbf {\bibinfo {volume} {53}},\ \bibinfo {pages} {9560} (\bibinfo
  {year} {2014})}\BibitemShut {NoStop}%
\bibitem [{\citenamefont {Long}\ \emph {et~al.}(2015)\citenamefont {Long},
  \citenamefont {Rankin},\ and\ \citenamefont {Ben-Amotz}}]{Long2015}%
  \BibitemOpen
  \bibfield  {author} {\bibinfo {author} {\bibfnamefont {J.~A.}\ \bibnamefont
  {Long}}, \bibinfo {author} {\bibfnamefont {B.~M.}\ \bibnamefont {Rankin}}, \
  and\ \bibinfo {author} {\bibfnamefont {D.}~\bibnamefont {Ben-Amotz}},\ }\href
  {\doibase 10.1021/jacs.5b06655} {\bibfield  {journal} {\bibinfo  {journal}
  {J. Am. Chem. Soc.}\ }\textbf {\bibinfo {volume} {137}},\ \bibinfo {pages}
  {10809} (\bibinfo {year} {2015})}\BibitemShut {NoStop}%
\bibitem [{\citenamefont {Pattenaude}\ \emph {et~al.}(2016)\citenamefont
  {Pattenaude}, \citenamefont {Rankin}, \citenamefont {Mochizuki},\ and\
  \citenamefont {Ben-Amotz}}]{Pattenaude2016}%
  \BibitemOpen
  \bibfield  {author} {\bibinfo {author} {\bibfnamefont {S.~R.}\ \bibnamefont
  {Pattenaude}}, \bibinfo {author} {\bibfnamefont {B.~M.}\ \bibnamefont
  {Rankin}}, \bibinfo {author} {\bibfnamefont {K.}~\bibnamefont {Mochizuki}}, \
  and\ \bibinfo {author} {\bibfnamefont {D.}~\bibnamefont {Ben-Amotz}},\ }\href
  {\doibase 10.1039/c6cp04379h} {\bibfield  {journal} {\bibinfo  {journal}
  {Phys. Chem. Chem. Phys.}\ }\textbf {\bibinfo {volume} {18}},\ \bibinfo
  {pages} {24937} (\bibinfo {year} {2016})}\BibitemShut {NoStop}%
\bibitem [{\citenamefont {Mochizuki}\ and\ \citenamefont
  {Ben-Amotz}(2017)}]{Mochizuki2017}%
  \BibitemOpen
  \bibfield  {author} {\bibinfo {author} {\bibfnamefont {K.}~\bibnamefont
  {Mochizuki}}\ and\ \bibinfo {author} {\bibfnamefont {D.}~\bibnamefont
  {Ben-Amotz}},\ }\href {\doibase 10.1021/acs.jpclett.7b00363} {\bibfield
  {journal} {\bibinfo  {journal} {J. Phys. Chem. Lett.}\ }\textbf {\bibinfo
  {volume} {8}},\ \bibinfo {pages} {1360} (\bibinfo {year} {2017})}\BibitemShut
  {NoStop}%
\bibitem [{\citenamefont {Kanno}\ \emph {et~al.}(1998)\citenamefont {Kanno},
  \citenamefont {Tomikawa},\ and\ \citenamefont {Mishima}}]{Kanno1998}%
  \BibitemOpen
  \bibfield  {author} {\bibinfo {author} {\bibfnamefont {H.}~\bibnamefont
  {Kanno}}, \bibinfo {author} {\bibfnamefont {K.}~\bibnamefont {Tomikawa}}, \
  and\ \bibinfo {author} {\bibfnamefont {O.}~\bibnamefont {Mishima}},\ }\href
  {\doibase 10.1016/S0009-2614(98)00813-6} {\bibfield  {journal} {\bibinfo
  {journal} {Chem. Phys. Lett.}\ }\textbf {\bibinfo {volume} {293}},\ \bibinfo
  {pages} {412} (\bibinfo {year} {1998})}\BibitemShut {NoStop}%
\bibitem [{\citenamefont {Takasu}\ \emph {et~al.}(2003)\citenamefont {Takasu},
  \citenamefont {Iwai},\ and\ \citenamefont {Nishio}}]{Takasu2003}%
  \BibitemOpen
  \bibfield  {author} {\bibinfo {author} {\bibfnamefont {Y.}~\bibnamefont
  {Takasu}}, \bibinfo {author} {\bibfnamefont {K.}~\bibnamefont {Iwai}}, \ and\
  \bibinfo {author} {\bibfnamefont {I.}~\bibnamefont {Nishio}},\ }\href
  {\doibase 10.1143/JPSJ.72.1287} {\bibfield  {journal} {\bibinfo  {journal}
  {J. Phys. Soc. Jpn.}\ }\textbf {\bibinfo {volume} {72}},\ \bibinfo {pages}
  {1287} (\bibinfo {year} {2003})}\BibitemShut {NoStop}%
\bibitem [{\citenamefont {Sugahara}\ \emph {et~al.}(2005)\citenamefont
  {Sugahara}, \citenamefont {Sugahara},\ and\ \citenamefont
  {Ohgaki}}]{Sugahara2005}%
  \BibitemOpen
  \bibfield  {author} {\bibinfo {author} {\bibfnamefont {K.}~\bibnamefont
  {Sugahara}}, \bibinfo {author} {\bibfnamefont {T.}~\bibnamefont {Sugahara}},
  \ and\ \bibinfo {author} {\bibfnamefont {K.}~\bibnamefont {Ohgaki}},\ }\href
  {\doibase 10.1021/je0496692} {\bibfield  {journal} {\bibinfo  {journal} {J.
  Chem. Eng. Data}\ }\textbf {\bibinfo {volume} {50}},\ \bibinfo {pages} {274}
  (\bibinfo {year} {2005})}\BibitemShut {NoStop}%
\bibitem [{\citenamefont {Chazallon}\ \emph {et~al.}(2007)\citenamefont
  {Chazallon}, \citenamefont {Focsa}, \citenamefont {Charlou}, \citenamefont
  {Bourry},\ and\ \citenamefont {Donval}}]{Chazallon2007}%
  \BibitemOpen
  \bibfield  {author} {\bibinfo {author} {\bibfnamefont {B.}~\bibnamefont
  {Chazallon}}, \bibinfo {author} {\bibfnamefont {C.}~\bibnamefont {Focsa}},
  \bibinfo {author} {\bibfnamefont {J.-L.}\ \bibnamefont {Charlou}}, \bibinfo
  {author} {\bibfnamefont {C.}~\bibnamefont {Bourry}}, \ and\ \bibinfo {author}
  {\bibfnamefont {J.-P.}\ \bibnamefont {Donval}},\ }\href {\doibase
  10.1016/j.chemgeo.2007.06.012} {\bibfield  {journal} {\bibinfo  {journal}
  {Chem. Geol.}\ }\textbf {\bibinfo {volume} {244}},\ \bibinfo {pages} {175}
  (\bibinfo {year} {2007})}\BibitemShut {NoStop}%
\bibitem [{\citenamefont {Chau}\ and\ \citenamefont
  {Hardwick}(1998)}]{Chau1998}%
  \BibitemOpen
  \bibfield  {author} {\bibinfo {author} {\bibfnamefont {P.}~\bibnamefont
  {Chau}}\ and\ \bibinfo {author} {\bibfnamefont {A.~J.}\ \bibnamefont
  {Hardwick}},\ }\href {\doibase 10.1080/002689798169195} {\bibfield  {journal}
  {\bibinfo  {journal} {Mol. Phys.}\ }\textbf {\bibinfo {volume} {93}},\
  \bibinfo {pages} {511} (\bibinfo {year} {1998})}\BibitemShut {NoStop}%
\bibitem [{\citenamefont {Errington}\ and\ \citenamefont
  {Debenedetti}(2001)}]{Errington2001a}%
  \BibitemOpen
  \bibfield  {author} {\bibinfo {author} {\bibfnamefont {J.}~\bibnamefont
  {Errington}}\ and\ \bibinfo {author} {\bibfnamefont {P.}~\bibnamefont
  {Debenedetti}},\ }\href {\doibase 10.1038/35053024} {\bibfield  {journal}
  {\bibinfo  {journal} {Nature}\ }\textbf {\bibinfo {volume} {409}},\ \bibinfo
  {pages} {318} (\bibinfo {year} {2001})}\BibitemShut {NoStop}%
\bibitem [{\citenamefont {Paolantoni}\ \emph {et~al.}(2009)\citenamefont
  {Paolantoni}, \citenamefont {Faginas~Lago}, \citenamefont {Alberti},\ and\
  \citenamefont {Lagan{\`{a}}}}]{Paolantoni2009}%
  \BibitemOpen
  \bibfield  {author} {\bibinfo {author} {\bibfnamefont {M.}~\bibnamefont
  {Paolantoni}}, \bibinfo {author} {\bibfnamefont {N.}~\bibnamefont
  {Faginas~Lago}}, \bibinfo {author} {\bibfnamefont {M.}~\bibnamefont
  {Alberti}}, \ and\ \bibinfo {author} {\bibfnamefont {A.}~\bibnamefont
  {Lagan{\`{a}}}},\ }\href {\doibase 10.1021/jp9052083} {\bibfield  {journal}
  {\bibinfo  {journal} {J. Phys. Chem. A}\ }\textbf {\bibinfo {volume} {113}},\
  \bibinfo {pages} {15100} (\bibinfo {year} {2009})}\BibitemShut {NoStop}%
\bibitem [{\citenamefont {Napoli}\ \emph {et~al.}(2017)\citenamefont {Napoli},
  \citenamefont {Marsalek},\ and\ \citenamefont {Markland}}]{Napoli2007}%
  \BibitemOpen
  \bibfield  {author} {\bibinfo {author} {\bibfnamefont {J.~A.}\ \bibnamefont
  {Napoli}}, \bibinfo {author} {\bibfnamefont {O.}~\bibnamefont {Marsalek}}, \
  and\ \bibinfo {author} {\bibfnamefont {T.~E.}\ \bibnamefont {Markland}},\
  }\href {http://arxiv.org/abs/1709.05740} {\bibfield  {journal} {\bibinfo
  {journal} {arXiv:1709.05740}\ } (\bibinfo {year} {2017})}\BibitemShut
  {NoStop}%
\end{thebibliography}%

\newpage



\section*{Supplementary material}

**SUPPORTING INFORMATION FOR:**

**The Interplay of Structure and Dynamics in the Raman Spectrum of Liquid Water over the Full Frequency and Temperature Range**

Tobias Morawietz,[a] Ondrej Marsalek,[b] Shannon R. Pattenaude,[c] Louis M. Streacker,[c] Dor Ben-Amotz,[c] and Thomas E. Markland[a,*]

a) Department of Chemistry, Stanford University, Stanford, CA 94305, United States

b) Faculty of Mathematics and Physics, Charles University, Ke Karlovu 3, 121 16 Prague 2, Czech Republic

c) Department of Chemistry, Purdue University, West Lafayette, IN 47907, United States

*Email: tmarkland@stanford.edu

# 1. Neural network parametrization

The parameters of the neural network potential (NNP)[1–3] employed for the molecular dynamics (MD) simulations were optimized with the neural network code RuNNer.[4] Previously, NNPs have been successfully applied to perform large-scale simulations of liquid water and ice based on a series of density functionals (RPBE, RPBE-D3, BLYP, BLYP-D3).[5] Here we parametrized a new potential based on the revPBE-D3 density functional. As in Ref. 5, the NNP consists of two atomic neural networks (describing the local environments of hydrogen and oxygen atoms, respectively), both with two hidden layers containing 25 nodes each. For the hidden layer nodes, hyperbolic tangent was used as the activation function, while a linear function was used for the output node. Atom-centered symmetry functions[6] of type 2 and 4 were used as descriptors. To more accurately describe the spectroscopic properties of liquid water, the symmetry function set used in Ref. 5 was modified and extended. The parameters of the extended symmetry function set are reported in Tables S1 and S2.

The NNP was trained to the energies and forces of about 10,000 condensed-phase configurations of liquid water, obtained from classical and path integral simulations at temperatures ranging from 260 K to 370 K employing ab initio molecular dynamics (AIMD)[7] and preliminary NNPs. The full reference set was split into a training set (90 % of all configurations) used to optimize the parameters and an independent validation set (10 % of all configurations). Root-mean-squared errors of energies and forces in the final NNP training (validation) set were 1.9 (2.0) meV/$H_2O$ and 72 (71) meV/Å, respectively. Figures S1 and S2 show the accuracy of the NNP with respect to the reference configurations for all energies and force components.

To further assess the accuracy of the NNP, in Figures S3 and S4 we show a comparison of structural (radial distribution functions) and spectroscopic (vibrational density of states (VDOS)) properties with the results of direct AIMD simulations employing the same density functional.[7] While the VDOS spectrum obtained from a different NNP based on the B3LYP-D3 density functional has been shown to display noticeable differences compared to AIMD simulations (especially in the bend region),[8] the revPBE-D3-based NNP reported here yields a VDOS that is almost indistinguishable from the AIMD results.

**Table S1.** Parameters $r_s$ (in Bohr), $r_{cut}$ (in Bohr), $\eta$ (in Bohr$^{-2}$), $\lambda$, and $\zeta$ of symmetry functions of type G2 (1-30) and type G4 (31-51) describing the local environment of hydrogen atoms.

| No. | Element j | Element k | $r_s$ | $r_{cut}$ | $\eta$ | $\lambda$ | $\zeta$ |
|---|---|---|---|---|---|---|---|
| 1 | O | - | 0.0 | 4 | 0.001 | - | - |
| 2 | O | - | 0.0 | 4 | 0.010 | - | - |
| 3 | O | - | 0.0 | 4 | 0.030 | - | - |
| 4 | O | - | 0.0 | 4 | 0.060 | - | - |
| 5 | O | - | -1.5 | 4 | 0.150 | - | - |
| 6 | O | - | 0.0 | 4 | 0.150 | - | - |
| 7 | O | - | -1.5 | 4 | 0.300 | - | - |
| 8 | O | - | -1.5 | 4 | 0.600 | - | - |
| 9 | H | - | 0.0 | 6 | 0.001 | - | - |
| 10 | H | - | 0.0 | 6 | 0.010 | - | - |
| 11 | H | - | 0.0 | 6 | 0.030 | - | - |
| 12 | H | - | 0.0 | 6 | 0.060 | - | - |
| 13 | H | - | -1.5 | 6 | 0.150 | - | - |
| 14 | H | - | 0.0 | 6 | 0.150 | - | - |
| 15 | H | - | -1.5 | 6 | 0.300 | - | - |
| 16 | H | - | 0.0 | 12 | 0.001 | - | - |
| 17 | O | - | 0.0 | 12 | 0.001 | - | - |
| 18 | H | - | 0.0 | 12 | 0.010 | - | - |
| 19 | O | - | 0.0 | 12 | 0.010 | - | - |
| 20 | H | - | 0.0 | 12 | 0.030 | - | - |
| 21 | O | - | 0.0 | 12 | 0.030 | - | - |
| 22 | H | - | 0.0 | 12 | 0.060 | - | - |
| 23 | O | - | 0.0 | 12 | 0.060 | - | - |
| 24 | H | - | -1.5 | 12 | 0.150 | - | - |
| 25 | O | - | -1.5 | 12 | 0.150 | - | - |
| 26 | H | - | 0.0 | 12 | 0.150 | - | - |
| 27 | O | - | 0.0 | 12 | 0.150 | - | - |
| 28 | H | - | -1.5 | 12 | 0.300 | - | - |
| 29 | O | - | -1.5 | 12 | 0.300 | - | - |
| 30 | O | - | -1.5 | 12 | 0.600 | - | - |
| 31 | H | O | 0.0 | 6 | 0.010 | -1 | 1 |
| 32 | H | O | 0.0 | 6 | 0.010 | 1 | 1 |
| 33 | H | O | 0.0 | 6 | 0.010 | -1 | 2 |
| 34 | H | O | 0.0 | 6 | 0.010 | 1 | 2 |
| 35 | H | O | 0.0 | 6 | 0.010 | 1 | 3 |
| 36 | H | O | 0.0 | 8 | 0.010 | -1 | 1 |
| 37 | H | O | 0.0 | 8 | 0.010 | 1 | 1 |
| 38 | H | O | 0.0 | 8 | 0.010 | -1 | 2 |
| 39 | H | O | 0.0 | 8 | 0.010 | 1 | 2 |
| 40 | H | O | 0.0 | 8 | 0.010 | 1 | 3 |
| 41 | O | O | 0.0 | 12 | 0.001 | -1 | 4 |
| 42 | O | O | 0.0 | 12 | 0.001 | 1 | 4 |
| 43 | H | O | 0.0 | 12 | 0.010 | -1 | 4 |
| 44 | H | O | 0.0 | 12 | 0.010 | 1 | 4 |
| 45 | H | O | 0.0 | 12 | 0.030 | -1 | 1 |
| 46 | O | O | 0.0 | 12 | 0.030 | -1 | 1 |
| 47 | H | O | 0.0 | 12 | 0.030 | 1 | 1 |
| 48 | O | O | 0.0 | 12 | 0.030 | 1 | 1 |
| 49 | H | O | 0.0 | 12 | 0.070 | -1 | 1 |
| 50 | H | O | 0.0 | 12 | 0.070 | 1 | 1 |
| 51 | H | O | 0.0 | 12 | 0.200 | 1 | 1 |

**Table S2.** Parameters $r_s$ (in Bohr), $r_{cut}$ (in Bohr), $\eta$ (in Bohr$^{-2}$), $\lambda$, and $\zeta$ of symmetry functions of type G2 (1-22) and type G4 (23-46) describing the local environment of oxygen atoms.

| No. | Element j | Element k | $r_s$ | $r_{cut}$ | $\eta$ | $\lambda$ | $\zeta$ |
|---|---|---|---|---|---|---|---|
| 1 | H | - | 0.0 | 4 | 0.001 | - | - |
| 2 | H | - | 0.0 | 4 | 0.010 | - | - |
| 3 | H | - | 0.0 | 4 | 0.030 | - | - |
| 4 | H | - | 0.0 | 4 | 0.060 | - | - |
| 5 | H | - | -1.5 | 4 | 0.150 | - | - |
| 6 | H | - | 0.0 | 4 | 0.150 | - | - |
| 7 | H | - | -1.5 | 4 | 0.300 | - | - |
| 8 | H | - | -1.5 | 4 | 0.600 | - | - |
| 9 | H | - | 0.0 | 12 | 0.001 | - | - |
| 10 | O | - | 0.0 | 12 | 0.001 | - | - |
| 11 | H | - | 0.0 | 12 | 0.010 | - | - |
| 12 | O | - | 0.0 | 12 | 0.010 | - | - |
| 13 | H | - | -1.5 | 12 | 0.030 | - | - |
| 14 | O | - | 0.0 | 12 | 0.030 | - | - |
| 15 | O | - | -1.5 | 12 | 0.060 | - | - |
| 16 | H | - | 0.0 | 12 | 0.060 | - | - |
| 17 | O | - | 0.0 | 12 | 0.060 | - | - |
| 18 | O | - | 0.0 | 12 | 0.090 | - | - |
| 19 | H | - | 0.0 | 12 | 0.150 | - | - |
| 20 | H | - | 0.0 | 12 | 0.150 | - | - |
| 21 | H | - | 0.0 | 12 | 0.300 | - | - |
| 22 | H | - | 0.0 | 12 | 0.600 | - | - |
| 23 | H | H | 0.0 | 6 | 0.010 | -1 | 1 |
| 24 | H | H | 0.0 | 6 | 0.010 | 1 | 1 |
| 25 | H | H | 0.0 | 6 | 0.010 | -1 | 2 |
| 26 | H | H | 0.0 | 6 | 0.010 | 1 | 2 |
| 27 | H | H | 0.0 | 6 | 0.010 | -1 | 3 |
| 28 | H | H | 0.0 | 8 | 0.010 | -1 | 1 |
| 29 | H | H | 0.0 | 8 | 0.010 | 1 | 1 |
| 30 | H | H | 0.0 | 8 | 0.010 | -1 | 2 |
| 31 | H | H | 0.0 | 8 | 0.010 | 1 | 2 |
| 32 | H | H | 0.0 | 8 | 0.010 | -1 | 3 |
| 33 | H | O | 0.0 | 12 | 0.001 | -1 | 4 |
| 34 | O | O | 0.0 | 12 | 0.001 | -1 | 4 |
| 35 | H | O | 0.0 | 12 | 0.001 | 1 | 4 |
| 36 | O | O | 0.0 | 12 | 0.001 | 1 | 4 |
| 37 | H | H | 0.0 | 12 | 0.010 | -1 | 4 |
| 38 | H | H | 0.0 | 12 | 0.010 | 1 | 4 |
| 39 | H | H | 0.0 | 12 | 0.030 | -1 | 1 |
| 40 | H | O | 0.0 | 12 | 0.030 | -1 | 1 |
| 41 | O | O | 0.0 | 12 | 0.030 | -1 | 1 |
| 42 | H | H | 0.0 | 12 | 0.030 | 1 | 1 |
| 43 | H | O | 0.0 | 12 | 0.030 | 1 | 1 |
| 44 | O | O | 0.0 | 12 | 0.030 | 1 | 1 |
| 45 | H | H | 0.0 | 12 | 0.070 | -1 | 1 |
| 46 | H | H | 0.0 | 12 | 0.070 | 1 | 1 |

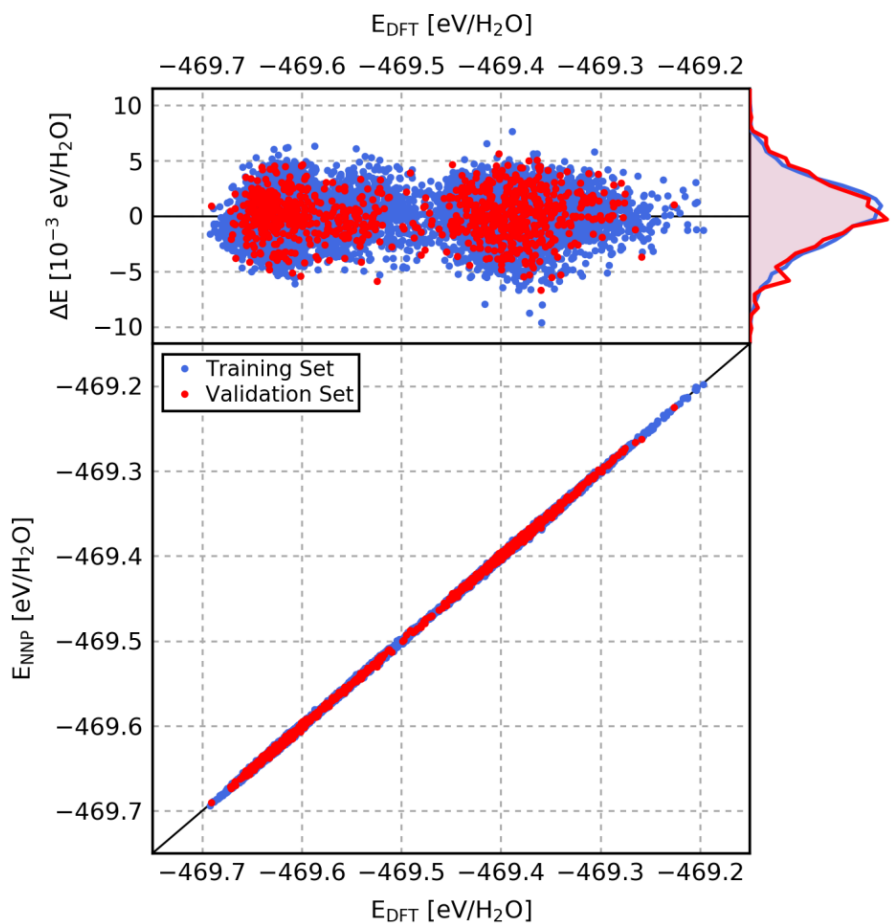

**Figure S1.** Comparison of energies obtained from the neural network potential (NNP) and density functional theory (DFT) for all configurations in the training and validation sets. The top panel shows the energy error ΔE = $E_{NNP}$ − $E_{DFT}$ as a function of the DFT energy together with the error distribution separately for the training and the validation set.

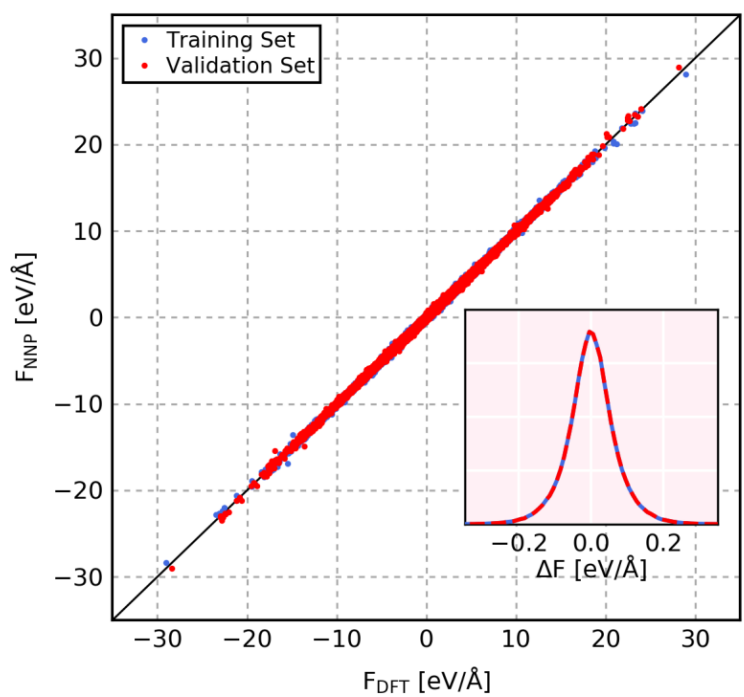

**Figure S2.** Comparison of Cartesian components of atomic forces obtained from the neural network potential (NNP) and density functional theory (DFT) for all configurations in the training and validation sets. The inset shows the distribution of the force error ΔF = $F_{NNP}$ − $F_{DFT}$, separately for the training and the validation sets.

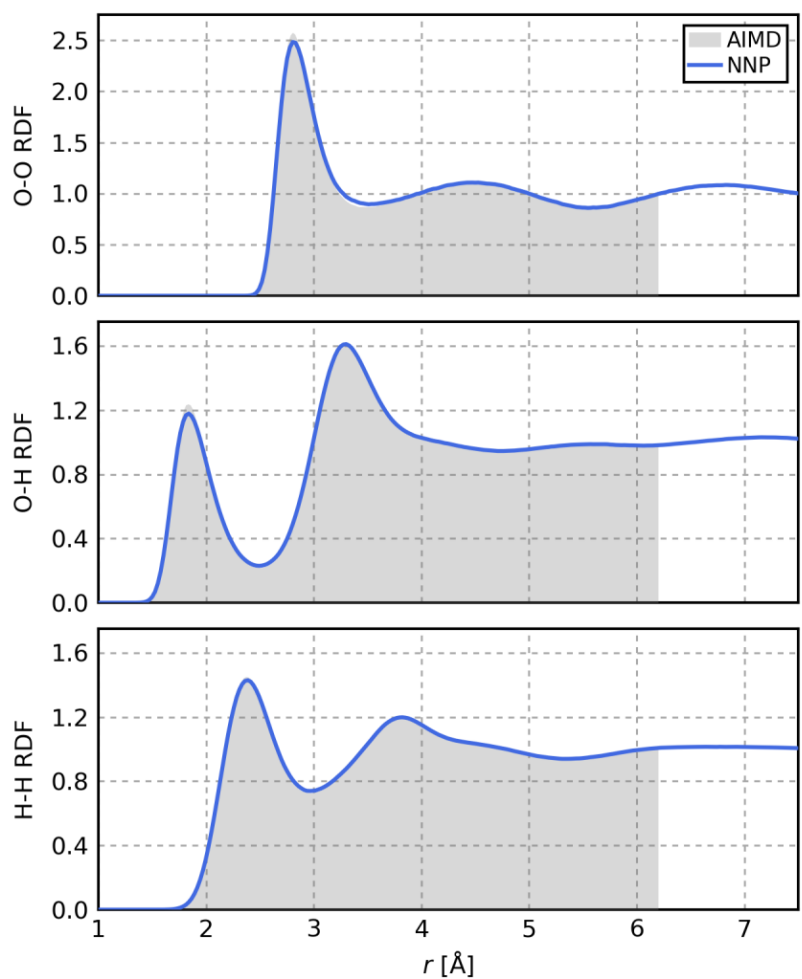

**Figure S3.** Oxygen-oxygen (O-O), oxygen-hydrogen (O-H), and hydrogen-hydrogen (H-H) radial distribution functions (RDFs) obtained from simulations with the neural network potential (NNP) at T = 300 K, compared to the results of direct AIMD simulations.[7] The RDFs were obtained from simulation cells containing 128 and 64 water molecules for the NNP and AIMD simulations, respectively.

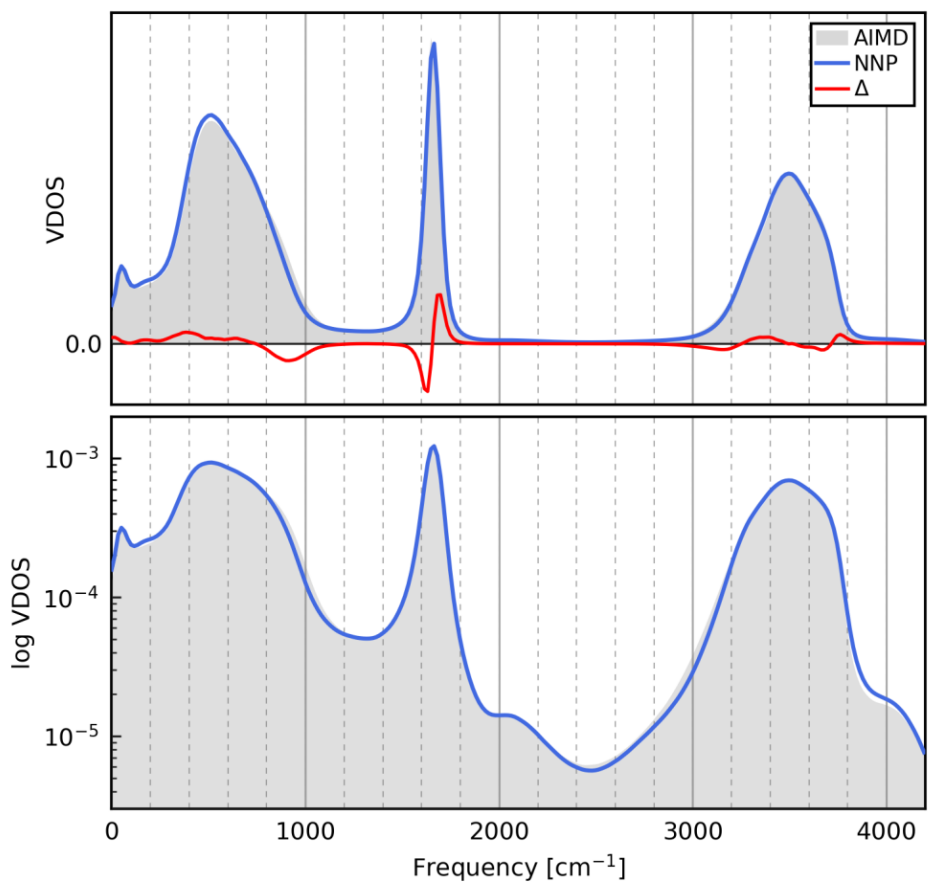

**Figure S4.** Vibrational density of states (VDOS) of the hydrogen atoms obtained from simulations with the neural network potential (NNP) at T = 300 K, compared to the results of direct AIMD simulations.[7] The top panel also shows the difference spectrum Δ = VDOS$_{NNP}$ − VDOS$_{DFT}$. The bottom panel shows both spectra on a logarithmic scale to highlight the low-intensity features.

## 2. Density functional theory calculations

Reference calculations were carried out with the CP2K program[9,10] using the revPBE exchange-correlation functional[11,12] with D3 dispersion correction[13] employing the same settings as those reported in Ref. 7. We have recently shown that AIMD simulations with classical nuclei and the particular density functional employed here yield water properties that are comparable to the results from path-integral simulations with the hybrid variant of this functional and generally in good agreement with experimental properties of water at ambient conditions.[7] The results that we present here confirm these observations and further show that the fortuitous error cancellation between deficiencies in this functional and the neglect of nuclear quantum effects appears to hold for the spectroscopic properties of liquid water over the whole liquid temperature range.

## 3. Simulation details

MD simulations were carried out with a modified version of the LAMMPS package,[5,14] for 1 ns per temperature with a time step of 0.5 fs. 200 ps were discarded for equilibration purposes, while the remaining 800 ps were used to compute the VDOS and the local tetrahedral order parameter, q. The stochastic velocity rescaling thermostat was used to sample the canonical ensemble.[15] Cubic simulation cells containing 128 water molecules with cell vectors corresponding to the experimental density at the given temperature were employed. The polarizability tensor of the system was obtained from density functional perturbation theory calculations[16] (employing the CP2K code) on configurations along trajectories obtained from the NNP simulations over an interval of 200 ps per temperature with a spacing of 2 fs.

## 4. Experimental Raman spectra

Raman spectra of liquid water were obtained as previously described,[17] except for the fact that the spectra were corrected to remove the wavelength dependent variations in the sensitivity of the CCD detector,[18–20] and converted to intensity units that are proportional to the number of Raman scattered photons per unit wavenumber (cm$^{-1}$). Additional experimental details will be provided in a subsequent publication that will contain the temperature-dependent spectra of both $H_2O$ and $D_2O$. Experimental isotropic and anisotropic spectra were constructed from the directly measured perpendicular (⊥) and parallel (∥) polarized spectra,

$$I(\omega)_{\text{iso}} = I(\omega)_{\parallel} - \tfrac{4}{3} I(\omega)_{\perp},$$

$$I(\omega)_{\text{aniso}} = \tfrac{4}{3} I(\omega)_\perp,$$

and shown in the reduced representation,[21,22]

$$R(\omega) = \frac{I(\omega) B(\omega) \omega}{\omega_0 (\omega_0 - \omega)^3},$$

where $\omega_0$ is the incident laser frequency and $B(\omega) = 1 - e^{-\hbar \omega \beta}$ is the Bose-Einstein correction factor, with $\beta = 1/(k_B T)$, and $\hbar = h/(2\pi)$.


### 5. Theoretical Raman spectra from autocorrelation functions

Reduced isotropic and anisotropic Raman spectra were obtained from autocorrelation functions of the corresponding polarizability tensor elements. Given a polarizability tensor $\boldsymbol{\alpha}(t)$, one can define its isotropic and anisotropic components as

$$\alpha(t) = \text{Tr}(\boldsymbol{\alpha}(t)),$$

$$\boldsymbol{\beta}(t) = \boldsymbol{\alpha}(t) - \mathbb{1}\,\text{Tr}(\boldsymbol{\alpha}(t))/3,$$

and their autocorrelation functions,

$$C_{\alpha\alpha}(t) = \langle \alpha(t_0) \alpha(t_0 + t) \rangle,$$

$$C_{\boldsymbol{\beta}\boldsymbol{\beta}}(t) = \langle \text{Tr}[\boldsymbol{\beta}(t_0) \boldsymbol{\beta}(t_0 + t)] \rangle.$$

In the following, $C(t)$ can be either $C_{\alpha\alpha}(t)$ or $C_{\boldsymbol{\beta}\boldsymbol{\beta}}(t)$, yielding isotropic or anisotropic intensities, respectively. Uncorrected Raman intensities, $I(\omega)$, are then obtained by

$$I(\omega) = \frac{\hbar \omega \beta}{(1 - e^{-\hbar \omega \beta})} \omega_0 (\omega_0 - \omega)^3 \int_{-\infty}^{\infty} e^{-i\omega t} C(t)\, dt,$$

where $\frac{\hbar \omega \beta}{(1-e^{-\hbar\omega\beta})}$ is the harmonic approximation quantum correction factor[16,23] and $\omega_0(\omega_0 - \omega)^3$ is the appropriate prefactor for comparisons with experimental Raman spectra obtained with photon counting detectors.[24] The theoretical Raman spectrum in its reduced representation $R(\omega)$ is then given by,

$$R(\omega) = \frac{I(\omega) B(\omega) \omega}{\omega_0 (\omega_0 - \omega)^3} = \hbar \omega^2 \beta \int_{-\infty}^{\infty} e^{-i\omega t} C(t)\, dt.$$

## 6. Hydrogen and Oxygen VDOS

Figure S5 shows the temperature dependence of the vibrational density of states (VDOS) plotted separately for the hydrogen and oxygen atoms on a linear scale.

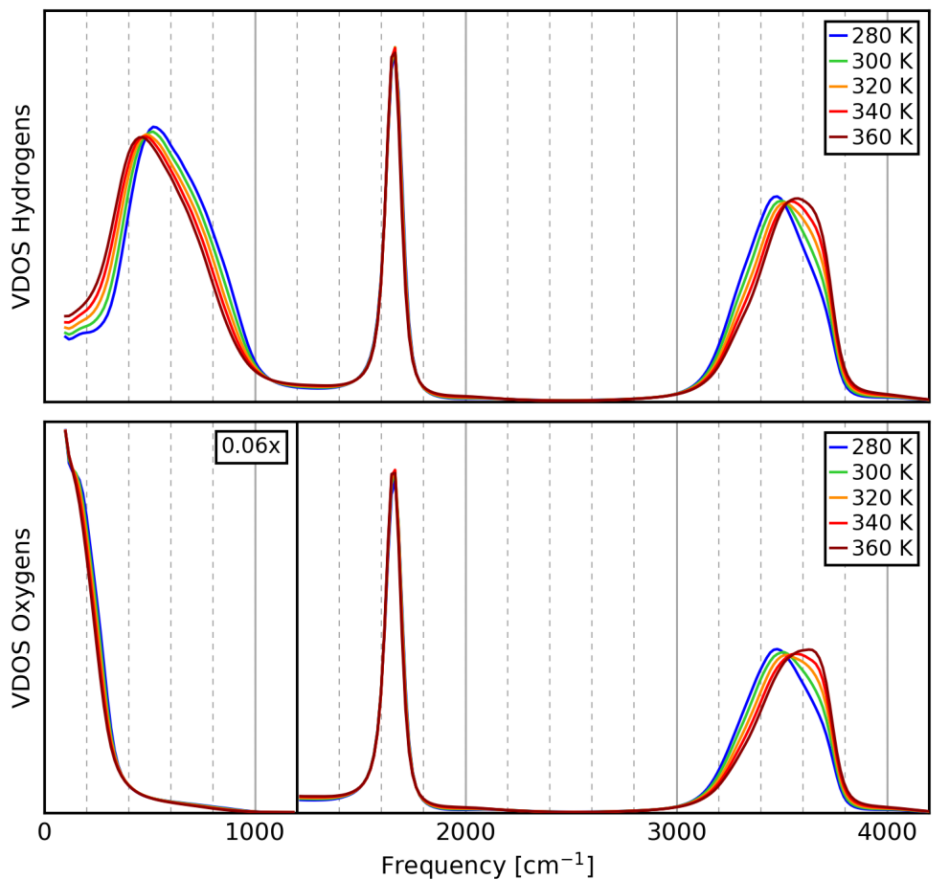

**Figure S5.** Temperature dependence of the vibrational density of states (VDOS) of hydrogen atoms (top) and oxygen atoms (bottom) obtained from simulations with the neural network potential. For clarity, the low-frequency region of the oxygen VDOS was scaled by a factor of 0.06.

## 7. Dependence of Time-dependent VDOS and two-dimensional frequency analysis on window width

In order to link vibrational motion to tetrahedral order, the time-dependent vibrational density of states (VDOS) was obtained as described in Ref. 7 employing a window width of 400 fs. The time-dependent VDOS was also used to obtain the synchronous two-dimensional correlation spectrum.[25] Figure S6 compares the correlation spectrum obtained with two different window

11

widths. From this one can observe that, while changing the window width slightly changes the shape of the cuts, it does not affect the correlations between the frequencies that were used to assign the combination bands at 2100 cm$^{-1}$ and 4100 cm$^{-1}$.

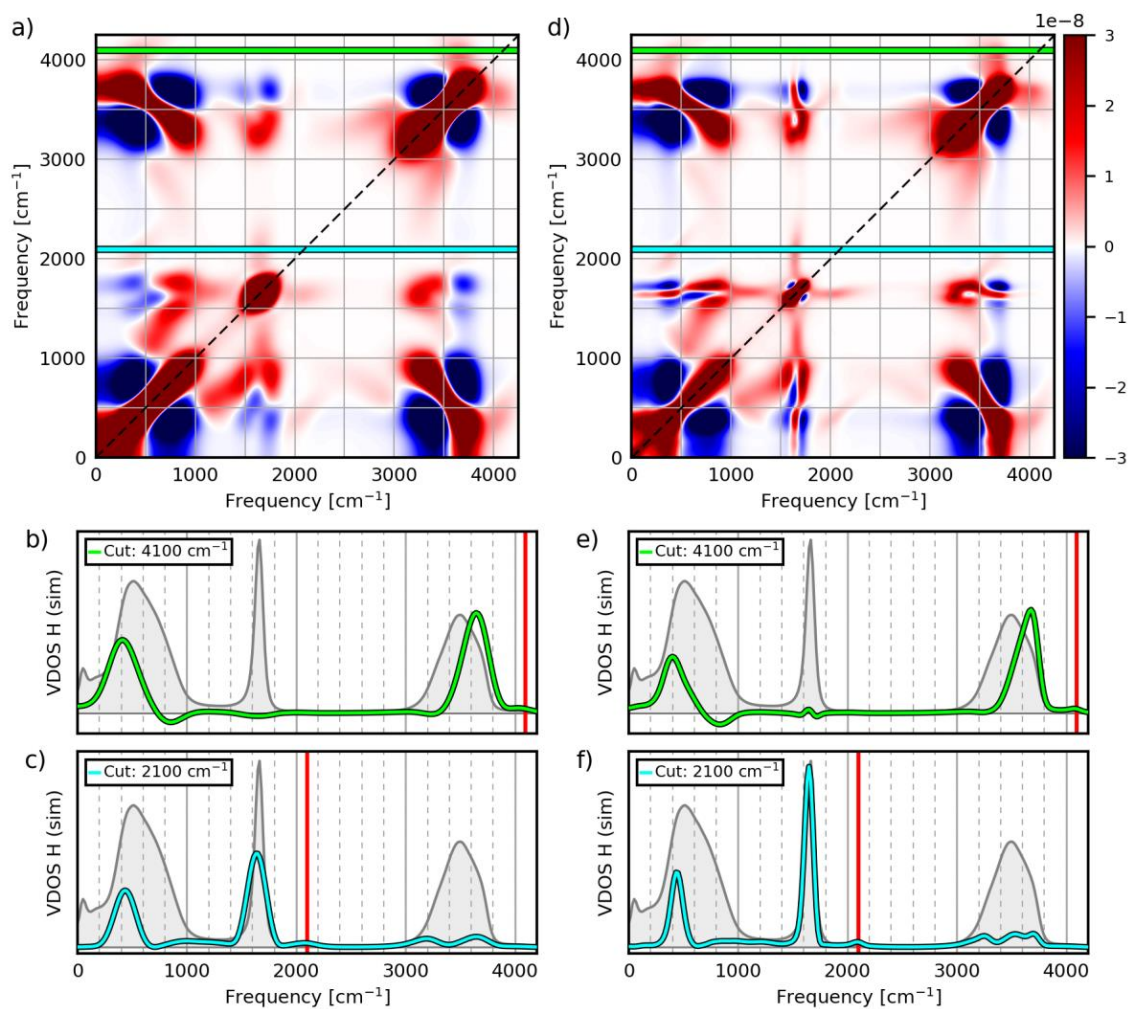

**Figure S6.** Comparison of the two-dimensional frequency analysis with two different window widths. Panels (a-c) show results obtained with a window width of 400 fs, whereas results in panels (d-f) where obtained with a window width of 1000 fs.

## 8. SMCR decomposition

Figure S7 shows the original temperature-dependent vibrational spectra and tetrahedral order distribution compared to the reconstructed data, obtained by combining the two temperature-independent SMCR components multiplied by their populations. Figure S8 shows the absolute difference between the original and the SMCR reconstructed data and reports the integrated fractional error,

12

$$\tfrac{1}{2}\int_0^\infty \left|I^{\text{full}}(\omega) - I^{\text{recon}}(\omega)\right| d\omega, \text{ and } \tfrac{1}{2}\int_0^\infty \left|P^{\text{full}}(q) - P^{\text{recon}}(q)\right| dq,$$

which varies between zero (perfect agreement) and one (no overlap) and is obtained by integrating over the absolute difference between the full data and the reconstructed data.

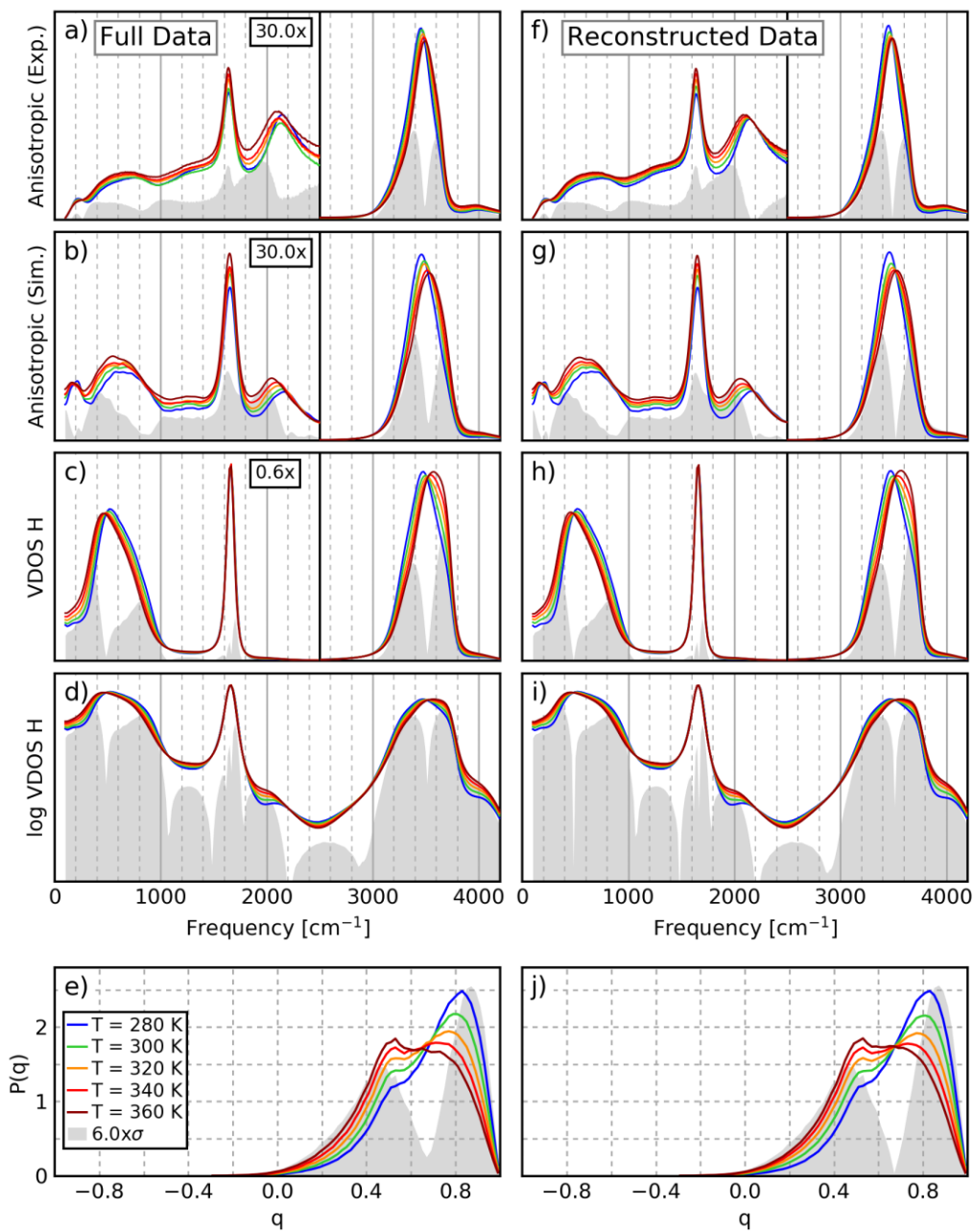

**Figure S7.** Comparison of the original temperature-dependent vibrational spectra and tetrahedral order parameter distribution (a-e) with the reconstructed data, obtained by combining the two temperature-independent SMCR components (f-j).

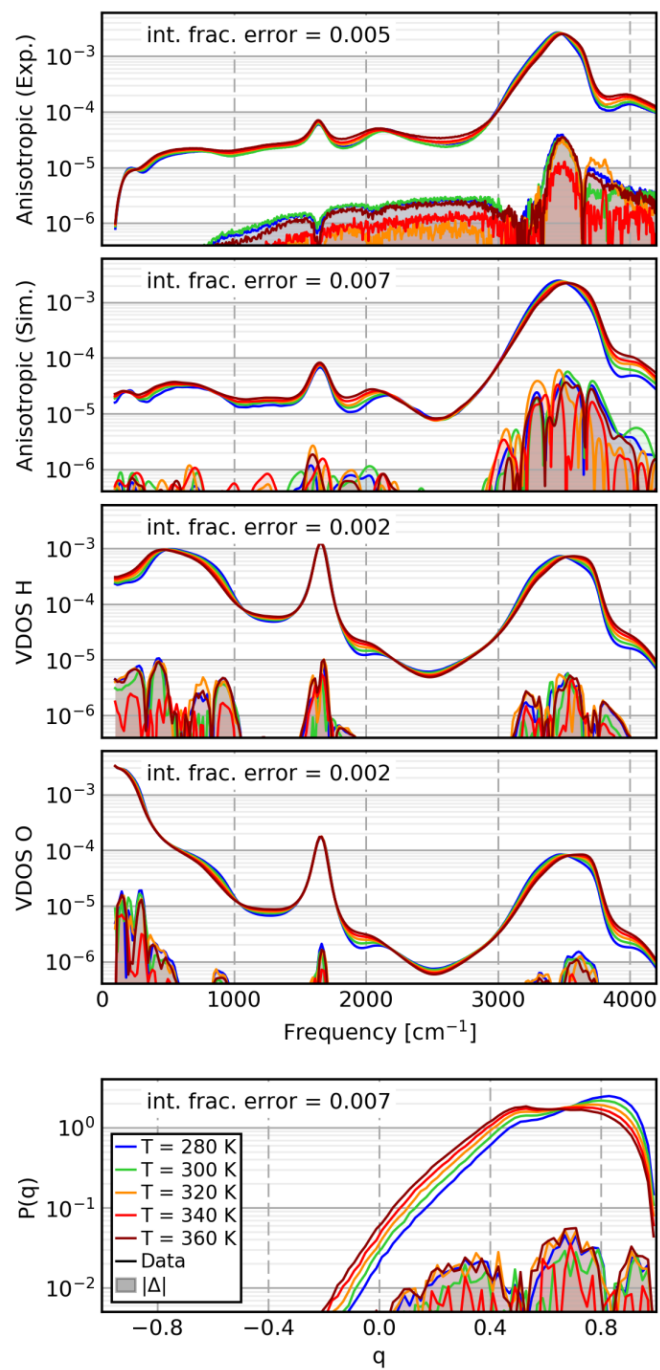

**Figure S8.** Temperature-dependent vibrational spectra and tetrahedral order parameter distribution (solid lines) compared to the absolute difference |Δ| (colored shading) between the original and the SMCR reconstructed data. All data sets are plotted on a logarithmic scale for a better comparison with the differences. Note that the error is consistently 1-2 orders of magnitude smaller than the original data.

14

**References**

(1)     Behler, J.; Parrinello, M. Generalized Neural-Network Representation of High-Dimensional Potential-Energy Surfaces. *Phys. Rev. Lett.* **2007**, *98* (14), 146401.

(2)     Artrith, N.; Urban, A. An Implementation of Artificial Neural-Network Potentials for Atomistic Materials Simulations: Performance for TiO2. *Comput. Mater. Sci.* **2016**, *114*, 135–150.

(3)     Behler, J. Perspective: Machine Learning Potentials for Atomistic Simulations. *J. Chem. Phys.* **2016**, *145* (17), 170901.

(4)     Behler, J. RuNNer, A Program for Constructing High-Dimensional Neural Network Potentials. Universität Göttingen 2017.

(5)     Morawietz, T.; Singraber, A.; Dellago, C.; Behler, J. How van Der Waals Interactions Determine the Unique Properties of Water. *Proc. Natl. Acad. Sci.* **2016**, *113* (30), 8368–8373.

(6)     Behler, J. Atom-Centered Symmetry Functions for Constructing High-Dimensional Neural Network Potentials. *J. Chem. Phys.* **2011**, *134*, 74106.

(7)     Marsalek, O.; Markland, T. E. Quantum Dynamics and Spectroscopy of Ab Initio Liquid Water: The Interplay of Nuclear and Electronic Quantum Effects. *J. Phys. Chem. Lett.* **2017**, *8* (7), 1545–1551.

(8)     Kapil, V.; Behler, J.; Ceriotti, M. High Order Path Integrals Made Easy. *J. Chem. Phys.* **2016**, *145* (23), 234103.

(9)     VandeVondele, J.; Krack, M.; Mohamed, F.; Parrinello, M.; Chassaing, T.; Hutter, J. Quickstep: Fast and Accurate Density Functional Calculations Using a Mixed Gaussian and Plane Waves Approach. *Comput. Phys. Commun.* **2005**, *167* (2), 103–128.

(10)    Hutter, J.; Iannuzzi, M.; Schiffmann, F.; VandeVondele, J. CP2K: Atomistic Simulations of Condensed Matter Systems. *Wiley Interdiscip. Rev. Comput. Mol. Sci.* **2014**, *4* (1), 15–25.

(11) Perdew, J. P.; Burke, K.; Ernzerhof, M. Generalized Gradient Approximation Made Simple. *Phys. Rev. Lett.* **1996**, *77* (18), 3865–3868.

(12) Zhang, Y.; Yang, W. Comment on "Generalized Gradient Approximation Made Simple." *Phys. Rev. Lett.* **1998**, *80* (4), 890–890.

(13) Grimme, S.; Antony, J.; Ehrlich, S.; Krieg, H. A Consistent and Accurate Ab Initio Parametrization of Density Functional Dispersion Correction (DFT-D) for the 94 Elements H-Pu. *J. Chem. Phys.* **2010**, *132* (15), 154104.

(14) Plimpton, S. Fast Parallel Algorithms for Short-Range Molecular Dynamics. *Journal of Computational Physics*. **1995**, 117, pp 1–19.

(15) Bussi, G.; Donadio, D.; Parrinello, M. Canonical Sampling through Velocity Rescaling. *J. Chem. Phys.* **2007**, *126*, 14101.

(16) Luber, S.; Iannuzzi, M.; Hutter, J. Raman Spectra from *Ab Initio* Molecular Dynamics and Its Application to Liquid *S*-Methyloxirane. *J. Chem. Phys.* **2014**, *141* (9), 94503.

(17) Davis, J. G.; Gierszal, K. P.; Wang, P.; Ben-Amotz, D. Water Structural Transformation at Molecular Hydrophobic Interfaces. *Nature* **2012**, *491* (7425), 582–585.

(18) Choquette, S. J.; Etz, E. S.; Hurst, W. S.; Blackburn, D. H.; Leigh, S. D. Relative Intensity Correction of Raman Spectrometers: NIST SRMs 2241 through 2243 for 785 Nm, 532 Nm, and 488 nm/514.5 Nm Excitation. *Appl. Spectrosc.* **2007**, *61* (2), 117–129.

(19) Drive, B. Standard Guide for Relative Intensity Correction of Raman Spectrometers. **2014**, *E2911*, 1–12.

(20) Ball, P. Chemical Physics: How to Keep Dry in Water. *Nature* **2003**, *423* (6935), 25–26.

(21) Brooker, M. H.; Nielsen, O. F.; Praestgaard, E. Assessment of Correction Procedures for Reduction of Raman Spectra. *J. Raman Spectrosc.* **1988**, *19* (2), 71–78.

(22) Brooker, M. H.; Hancock, G.; Rice, B. C.; Shapter, J. Raman Frequency and Intensity Studies of Liquid $H_2O$, $H_2^{18}O$ and $D_2O$. *J. Raman Spectrosc.* **1989**, *20* (10), 683–694.

(23) Ramirez, R.; Lopez-Ciudad, T.; Kumar P, P.; Marx, D. Quantum Corrections to Classical

Time-Correlation Functions: Hydrogen Bonding and Anharmonic Floppy Modes. *J. Chem. Phys.* **2004**, *121* (9), 3973–3983.

(24) McCreery, R. L. *Raman Spectroscopy for Chemical Analysis*; John Wiley & Sons, Inc.: New York, 2000.

(25) Noda, I. Generalized Two-Dimensional Correlation Method Applicable to Infrared, Raman, and Other Types of Spectroscopy. *Appl. Spectrosc.* **1993**, *47* (9), 1329–1336.



\end{document}